\documentclass{mcom-l}

\usepackage{amssymb}

\usepackage{graphicx}
\usepackage{threeparttable}
\usepackage{algorithmicx,algorithm}
\usepackage{algpseudocode}

\theoremstyle{definition}
\newtheorem{sec2lemma1x}{Lemma}[section]
\newtheorem{sec2lemma2x}[sec2lemma1x]{Lemma}
\newtheorem{sec2lemma3x}[sec2lemma1x]{Lemma}
\newtheorem{sec2lemma4x}[sec2lemma1x]{Lemma}

\newtheorem{sec3_remark1}{Remark}[section]
\newtheorem{sec3_remark2}[sec3_remark1]{Remark}
\newtheorem{sec3_remark3}[sec3_remark1]{Remark}

\newtheorem{sec3_B_remark1}[sec3_remark1]{Remark}
\newtheorem{sec3_B_remark2}[sec3_remark1]{Remark}
\newtheorem{sec3_B_remark3}[sec3_remark1]{Remark}
\newtheorem{sec3_B_remark4}[sec3_remark1]{Remark}

\newtheorem{sec3Cremark1}[sec3_remark1]{Remark}
\newtheorem{sec3Cremark2}[sec3_remark1]{Remark}
\newtheorem{sec3Cremark3}[sec3_remark1]{Remark}
\newtheorem{sec3Cremark4}[sec3_remark1]{Remark}

\newtheorem{sec3thm1}{Theorem}[section]
\newtheorem{sec3thm1x}[sec3thm1]{Theorem}
\newtheorem{sec3thm1xx}[sec3thm1]{Theorem}
\newtheorem{sec3thm2}[sec3thm1]{Theorem}
\newtheorem{sec3thm2aux1}[sec3thm1]{Theorem}
\newtheorem{sec3thm2aux2}[sec3thm1]{Theorem}

\newtheorem{sec3thm2x}[sec3thm1]{Theorem}

\newtheorem{sec3CThm01}[sec3thm1]{Theorem}
\newtheorem{sec3CThm02}[sec3thm1]{Theorem}
\newtheorem{sec3CThm03}[sec3thm1]{Theorem}
\newtheorem{sec3CThm04}[sec3thm1]{Theorem}
\newtheorem{sec3CThm05}[sec3thm1]{Theorem}

\newtheorem{sec3Bcor1}{Corollary}[section]
\newtheorem{sec3Bcor2}[sec3Bcor1]{Corollary}
\newtheorem{sec3Bcor3}[sec3Bcor1]{Corollary}
\newtheorem{sec3Bcor4}[sec3Bcor1]{Corollary}

\newtheorem{sec3Ccor1}[sec3Bcor1]{Corollary}
\newtheorem{sec3Ccor2}[sec3Bcor1]{Corollary}
\newtheorem{sec3Ccor3}[sec3Bcor1]{Corollary}
\newtheorem{sec3Ccor4}[sec3Bcor1]{Corollary}
\newtheorem{sec3Ccor5}[sec3Bcor1]{Corollary}

\newtheorem{sec3exp1}{Example}[section]

\newtheorem{sec3Bexp1}[sec3exp1]{Example}

\newtheorem{sec3lemma1}{Lemma}[section]
\newtheorem{sec3lemma2}[sec3lemma1]{Lemma}
\newtheorem{sec3lemma3}[sec3lemma1]{Lemma}
\newtheorem{sec3lemma4}[sec3lemma1]{Lemma}
\newtheorem{sec3lemma5}[sec3lemma1]{Lemma}
\newtheorem{sec3lemma6}[sec3lemma1]{Lemma}
\newtheorem{sec3lemma7}[sec3lemma1]{Lemma}
\newtheorem{sec3lemma8}[sec3lemma1]{Lemma}
\newtheorem{sec3lemma9}[sec3lemma1]{Lemma}
\newtheorem{sec3lemma10}[sec3lemma1]{Lemma}

\numberwithin{equation}{section}
\begin{document}

\title[Linear Complexity of Generalized Cyclotomic Quaternary Sequences ]{On the Linear Complexity of Generalized Cyclotomic Quaternary Sequences with Length $ 2pq $}

\author{Minglong Qi}
\address{School of Computer Science and Technology, Wuhan University of Technology, Mafangshan West Campus, 430070 Wuhan City, China}
\curraddr{}
\email{mlqiecully@163.com}

\author{Shengwu Xiong}
\address{}
\curraddr{}
\email{}

\author{Jingling Yuan}
\address{}
\curraddr{}
\email{}
\thanks{}

\author{Wenbi Rao}
\address{}
\curraddr{}
\email{}
\thanks{}

\author{Luo Zhong}
\address{}
\curraddr{}
\email{}
\thanks{}


%


\begin{abstract}
In this paper, the linear complexity over $ \mathbf{GF}(r) $   of   generalized cyclotomic quaternary sequences with period $ 2pq $  is determined, where $ r $ is an odd prime such that $ r \ge 5 $ and $ r\notin \lbrace p,q\rbrace $. The minimal value of the linear complexity  is equal to $ \tfrac{5pq+p+q+1}{4} $ which is greater than  the half of the period $ 2pq $. According to the Berlekamp-Massey algorithm, these sequences are viewed as enough good for the use in cryptography. We show also that if the character of the extension field $ \mathbf{GF}(r^{m}) $, $ r $, is chosen so that  $\bigl(\tfrac{r}{p}\bigr)=\bigl(\tfrac{r}{q}\bigr)=-1 $, $ r\nmid 3pq-1 $, and $ r\nmid 2pq-4 $, then the linear complexity can reach the maximal value equal to the length of the sequences.
\end{abstract}
\maketitle
\section{Introduction}
\label{sec1}
Pseudo-random sequences play important roles in fields such as communication systems, simulation, and cryptography \cite{B1,B2}. In  cryptography, sequences with good balance and high linear complexity are preferable\cite{B2,B3,B4}. According to the Berlekamp-Massey algorithm\cite{B5}, if the linear complexity is greater than the half of the period, then the sequences are viewed as enough good for cryptographic uses. Because of their  good algebraic structure\cite{B6,B8}, cyclotomic sequences of different periods and orders find many applications in cryptography and communication \cite{B4,B7,B8,B9,B10}.

A family of generalized cyclotomic quaternary sequences of length $ 2p $ was constructed in \cite{B11}, of which the linear complexity and the autocorrelation were studied in \cite{B10} and in \cite{B12}, respectively. In \cite{B13}, Chang et al  considered the linear complexity of quaternary cyclotomic sequences with period $ 2pq $.  Using the technique of Fourier spectrums of  sequences, the authors of \cite{B14} determined the linear complexity of generalized cyclotomic sequences with period $ pq $. The linear complexity of Whiteman\textquoteright s generalized cyclotomic binary sequences with the period $p^{m+1}q^{n+1}$ was computed out in \cite{B20}. Very recently in \cite[June 2015]{B21}, Li et al studied the linear complexity of generalized cyclotomic binary sequences with the length $2p^{m+1}q^{n+1}$. Because of easy and efficient hardware implementation, researches on  linear complexity and autocorrelation of generalized cyclotomic binary and quaternary sequences are intensive \cite{B7,B8,B9,B15,B16,B17}.

 In this paper, we consider the linear complexity over $ GF(r) $ of the generalized cyclotomic quaternary sequences with period $ 2pq $ constructed in \cite{B13}. The rest of the paper is organized as follows: in Section 2, basic definitions, notations  and related  lemmas of previous works needed to prove the main results are given. In Section 3, we give main theorems of this paper and their proofs. In Section 4, we give a brief conclusion and remarks.
 
 \section{Preliminaries}
 Let $ p,\,q $ be two distinct odd primes, $ r $ be an odd prime such that $ r \geq 5 $ and $ r \ne p,\,q $,  and $ m $ be the order of $ r $ modulo $ pq $. Then, by the Chinese Remainder Theorem, there is a $ 2pq^{th} $ root of unity in the extension field $ \mathbf{GF}(r^{m}) $. Throughout the paper, we keep the meanings of $ r $, $ m $, and $ \mathbf{GF}(r^{m}) $, unchanged. From here and hereafter, the meanings of the following symbols are kept unchanged: $ p,\,q $ are not only two distinct odd primes, but also $ \gcd(p-1,q-1)=2 $, which implicates that, the generalized cyclotomic quaternary sequences with the period $ 2pq $ is of order two. Let $ T=pq $, and $ \beta $ be a $ 2pq^{th} $ root of unity in $ \mathbf{GF}(r^{m}) $. We use $ \beta_{T} $, $ \beta_{p} $, $ \beta_{q} $, and $ \beta_{2} $, to denote a $ pq^{th} $, a $ p^{th} $, a $ q^{th} $ and a second root of unity in $ \mathbf{GF}(r^{m}) $, respectively. We define $ M_{p}=q\cdot(q^{-1}\bmod{p}) \pmod{pq}$, $ M_{2}=u\cdot(u^{-1}\bmod 2) \pmod {2u}$ where $ u\in \lbrace p,q,T\rbrace $, $ M_{q}=p\cdot(p^{-1}\bmod{q}) \pmod{pq}$, $ 2^{-1}_{p} =2^{-1}\pmod p$, $ 2^{-1}_{q} =2^{-1}\pmod q$, and $ 2^{-1}_{T} =2^{-1}\pmod T$. It is clear that, $ M_{2}\equiv u \pmod{2u} $, with $ u\in \lbrace p,q,T\rbrace $. In addition, let $ g $ be a common odd primitive root of $ p $ and $ q $, and $ x $ be an integer which is computed by the following congruent system:
 \begin{equation}\label{sec2_eq_x1}
 \begin{cases}
 x\equiv g\quad &\pmod{2p},\\
 x\equiv 1\quad &\pmod{2q}.
 \end{cases}
 \end{equation}
 If $ sq+tp=1 $, then by the Generalized Chinese Remainder Theorem, 
 \begin{equation}\label{sec2_eq_x2}
 x\equiv g\cdot s\cdot q+t\cdot p\pmod{2pq}.
 \end{equation}
 
 We define the following sets:
 \begin{equation}\label{sec2_eq_sets_1}
 \begin{cases}
 D_{0}^{(2T)}=<g> &\pmod {2T},\\
 D_{1}^{(2T)}=<g>x &\pmod {2T},\\
 D_{0}^{(T)}=<g> &\pmod {T},\\
 D_{1}^{(T)}=<g>x &\pmod {T},\\
 D_{0}^{(2p)}=<g^{2}> &\pmod {2p},\\
 D_{1}^{(2p)}=<g^{2}>g &\pmod {2p},\\
 D_{0}^{(2q)}=<g^{2}> &\pmod {2q},\\
 D_{1}^{(2q)}=<g^{2}>g &\pmod {2q},\\
 D_{0}^{(p)}=<g^{2}> &\pmod {p},\\
 D_{1}^{(p)}=<g^{2}>g &\pmod {p},\\
 D_{0}^{(q)}=<g^{2}> &\pmod {q},\\
 D_{1}^{(q)}=<g^{2}>g &\pmod {q}.
 \end{cases}
 \end{equation}
 From (\ref{sec2_eq_sets_1}), the following partitions are straightforward:
 \begin{equation}\label{sec2_eq_sets_2}
 \begin{cases}
 &Z^{*}_{2T}=D_{0}^{(2T)}\cup D_{1}^{(2T)},\ Z^{*}_{T}=D_{0}^{(T)}\cup D_{1}^{(T)},\\
 &Z^{*}_{2p}=D_{0}^{(2p)}\cup D_{1}^{(2p)},\ Z^{*}_{2q}=D_{0}^{(2q)}\cup D_{1}^{(2q)},\\
 &Z^{*}_{p}=D_{0}^{(p)}\cup D_{1}^{(p)},\ 
 Z^{*}_{q}=D_{0}^{(q)}\cup D_{1}^{(q)},\\
 &Z_{2T}=Z^{*}_{2T}\cup 2Z^{*}_{T}\cup 
 qZ^{*}_{2p}\cup pZ^{*}_{2q}
 \cup 2qZ^{*}_{p}\cup\\ 
 &2pZ^{*}_{q}
 \cup \lbrace pq\rbrace \cup \lbrace 0\rbrace.
 \end{cases}
 \end{equation}
 
 Define four parameters as follows:
 \begin{equation}\label{sec2_eq_parameters}
 \begin{cases}
 &A_{0}= \sum_{i\in D_{0}^{(p)}}\beta^{i}_{p},
 \ A_{1}= \sum_{i\in D_{1}^{(p)}}\beta^{i}_{p};\\
 &B_{0}= \sum_{i\in D_{0}^{(q)}}\beta^{i}_{q},
 \ B_{1}= \sum_{i\in D_{1}^{(q)}}\beta^{i}_{q}.
 \end{cases}
 \end{equation}
 
 Below is the definition of the generalized cyclotomic quaternary sequences of length $ 2T $ constructed by Chang et al in \cite{B13}, $ \left\{ s(t):t\pmod{2T}=0,1,\cdots,2T-1\right\} $ .
   \begin{equation}\label{lab_seq_chang}
   s(t)=
   \begin{cases}
   0 &\text{if}\  t\in \lbrace 0\rbrace,\\
   e &\text{if}\  t\in \lbrace T\rbrace,\\
   a &\text{if}\  t\in D_{0}^{(2T)}\cup qD_{0}^{(2p)}\cup pD_{0}^{(2q)},\\
   b\quad &\text{if}\  t\in D_{1}^{(2T)}\cup qD_{1}^{(2p)}\cup pD_{1}^{(2q)},\\
   c &\text{if}\  t\in 2D_{0}^{(T)}\cup 2qD_{0}^{(p)}\cup 2pD_{0}^{(q)},\\
   d &\text{if}\  t\in 2D_{1}^{(T)}\cup 2qD_{1}^{(p)}\cup 2pD_{1}^{(q)}.
   \end{cases}
   \end{equation}
   In (\ref{lab_seq_chang}), $ a,b,c$ and $d\in \mathbb{F}_{4} $, but $ e\in   \mathbb{F}_{4}^{*} $. Let $ \mathbb{S} =\lbrace 0,1,2,3\rbrace \subset \mathbf{GF}(r)$. Then, there exists always a  mapping from $ \mathbb{F}_{4} $ to $ \mathbb{S}  $, $ \psi $, such that $ \psi(a)=0, \psi(b)=1,\psi(c)=\psi(e)=2$ and $ \psi(d)=3 $.  We thus obtain a particular instance of (\ref{lab_seq_chang}):
   \begin{equation}\label{lab_seq_chang_ex}
     s(t)=
     \begin{cases}
     0 &\text{if}\  t\in \lbrace 0\rbrace,\\
     2 &\text{if}\  t\in \lbrace T\rbrace,\\
     0 &\text{if}\  t\in D_{0}^{(2T)}\cup qD_{0}^{(2p)}\cup pD_{0}^{(2q)},\\
     1\quad &\text{if}\  t\in D_{1}^{(2T)}\cup qD_{1}^{(2p)}\cup pD_{1}^{(2q)},\\
     2 &\text{if}\  t\in 2D_{0}^{(T)}\cup 2qD_{0}^{(p)}\cup 2pD_{0}^{(q)},\\
     3 &\text{if}\  t\in 2D_{1}^{(T)}\cup 2qD_{1}^{(p)}\cup 2pD_{1}^{(q)}.
     \end{cases}
     \end{equation}
    In this paper, we study the linear complexity of the sequences defined in (\ref{lab_seq_chang_ex}) over $ \mathbf{GF}(r^{m}) $.
    
     Let $ \{s(t):t=0,1,\cdots,N-1\} $ be a sequence over $\mathbf{GF}(r^{m}) $ with period $ N $. Then, the linear complexity of $ s(t) $ over $ \mathbf{GF}(r^{m}) $, denoted by $ LC(s) $, is the least integer $ L $ which satisfies next recurrent relation:
      \begin{equation*}
      s(t+L)=c_{L-1}s(t+L-1)+\cdots+c_{1}s(t+1)+c_{0}s(t)
      \end{equation*}
    for $ t\ge 0 $, where $ c_{0},c_{1},\cdots,c_{L-1} \in  \mathbf{GF}(r^{m}) $.
    The minimal polynomial and the generating polynomial over  $ \mathbf{GF}(r^{m})[x]  $
    related to $ s(t) $ are given below respectively
    \begin{equation}\label{lab_eq_GCD_GP_ex}
    \begin{split}
    M_{s}(x)&=x^{L}-\sum_{i=0}^{L-1}c_{i}x^{i},\\
    G_{s}(x)&=\sum_{t=0}^{N-1}s(t)x^{t}.
    \end{split}
    \end{equation}
    The following equation relates both the minimal polynomial and the generating polynomial of the sequence $ s(t) $:
    \begin{equation}\label{lab_eq_GCD_MiniPoly_ex}
    M_{s}(x)=\frac{x^{N}-1}{\gcd(x^{N}-1,G_{s}(x))}.
    \end{equation}
    From (\ref{lab_eq_GCD_MiniPoly_ex}), the linear complexity can be calculated by
    \begin{equation}\label{lab_eq_GCD_LC_eq_ex}
    LC(s)=\deg(M_{s}(x))=N-\deg(\gcd(x^{N}-1,G_{s}(x))).
    \end{equation}
    
    In this paper, we focus on how to compute the degree of $ \gcd $ in (\ref{lab_eq_GCD_LC_eq_ex}) over the extension field $ \mathbf{GF}(r^{m}) $,  for the sequence defined in (\ref{lab_seq_chang_ex}). We will compute out the number of zeros of the generating polynomial $ G_{s}(x)$  under the form $ \beta^{k} $ for $  k\in Z_{2T} $, from which  the degree of $ \gcd(x^{2T}-1,G_{s}(x)) $ can be deduced.
    
    We write down a series of lemmas of previous works related to and necessary for proof of the main theorems.
    \begin{sec2lemma1x}\label{lab_sec2_lemma1x}Let $ A_{0},A_{1},B_{0} $, and $ B_{1} $ be the parameters defined in (\ref{sec2_eq_parameters}).
     \begin{enumerate} 
     \item If $ p\equiv 1 \pmod 4 $, and $ q\equiv 1 \pmod 4 $, then $ A_{0}(1+A_{0}) \equiv \frac{p-1}{4} \pmod r$, and $ B_{0}(1+B_{0}) \equiv \frac{q-1}{4} \pmod r$.
     \item If $ p\equiv 3 \pmod 4 $, and $ q\equiv 3 \pmod 4 $, then $ A_{1}(1+A_{1}) \equiv -\frac{p+1}{4} \pmod r$, and $ B_{1}(1+B_{1}) \equiv -\frac{q+1}{4} \pmod r$.
     \end{enumerate}
     \end{sec2lemma1x}
     \begin{proof}
     Since $ r $ is an odd prime such that $ r\geq 5 $ and $ r\ne p,q $, and $ m $ is the order of $ r $ modulo $ T $, hence $ r^{m}\equiv 1\pmod{ pq} $. In other words, $ \gcd(r^{m},p)=1 $ and $ \gcd(r^{m},q)=1 $. In addition, $ \beta_{p} $ and $ \beta_{q} $ are the $ p^{th} $ root of unity and the $ q^{th} $ root of unity over the extension field $ \mathbf{GF}(r^{m}) $, respectively. From    (\ref{sec2_eq_parameters}), $  A_{0}= \sum_{i\in D_{0}^{(p)}}\beta^{i}_{p} $, and $ B_{0}= \sum_{i\in D_{0}^{(q)}}\beta^{i}_{q} $. Lemma \ref{lab_sec2_lemma1x} (1) follows from (13) in \cite{B4}. The proof of Lemma \ref{lab_sec2_lemma1x} (2) is similar to that of Lemma \ref{lab_sec2_lemma1x} (1), and mentioned at first time in \cite{B18}.
     \end{proof}
    \begin{sec2lemma2x}\cite{B10,B13}\label{lab_sec2_lemma2x}
    \begin{enumerate} 
    \item if $ a\in D_{i}^{(u)} $, then $ aD_{j}^{(u)}\pmod u = D_{i+j \pmod 2}^{(u)} $.
    \item $ D_{i}^{(2u)}\pmod u = D_{i}^{(u)} $.
    \end{enumerate}
    \end{sec2lemma2x} 
    Where $ u\in\lbrace p,q,T\rbrace $. 
    \begin{sec2lemma3x}\cite{B19}\label{lab_sec2_lemma3x}
     \begin{enumerate} 
     \item 
    $ 2\in D_{0}^{(u)} $ if only if $ u\equiv\pm 1\pmod 8  $, and $ 2\in D_{1}^{(u)} $ if only if $ u\equiv\pm 3\pmod 8  $, where $ u\in\lbrace p,q\rbrace $.
     \item
     (Quadratic Reciprocity).
     \begin{equation*}
     \left(\frac{q}{p}\right)= (-1)^{\frac{p-1}{2}\cdot \frac{q-1}{2}}\left(\frac{p}{q}\right).
     \end{equation*}
    \end{enumerate}
    \end{sec2lemma3x}
    \begin{sec2lemma4x}\cite{B14}\label{lab_sec2_lemma4x}
     $ A_{0} $ and $ A_{1} \in \mathbf{GF}(r)$ if only if $ r\in  D_{0}^{(p)}$.
    \end{sec2lemma4x}
   \section{Computing the linear complexity according to various cases}
         For easy expression of some results, we give the definition of the sets which classify the pairs of $ (p,q) $. Let $ s=p\pmod 8 $ and $ t=q\pmod 8 $, then
         \begin{equation*}
         \begin{split}
         (\pm{1},\pm{1})=\left\lbrace(p,q):s=\pm 1 \ \text{and}\ t=\pm 1\right\rbrace,\\
         (\pm{3},\pm{3})=\left\lbrace(p,q):s=\pm 3 \ \text{and}\ t=\pm 3\right\rbrace,\\
         (\pm{1},\pm{3})=\left\lbrace(p,q):s=\pm 1 \ \text{and}\ t=\pm 3\right\rbrace,\\
         (\pm{3},\pm{1})=\left\lbrace(p,q):s=\pm 3 \ \text{and}\ t=\pm 1\right\rbrace.
         \end{split}
         \end{equation*}
         As we require that $ \gcd(p-1,q-1)=2 $, it is clear that $ (p,q)\notin(1,1)\cup (-3,-3) \cup (1,-3)\cup (-3,1)$.
         
         Let $\left(\tfrac{r}{p}\right)  $ denote the Legendre symbol of $ r $ modulo $ p $. Next, we give a series of lemmas necessary to prove the  main theorems.
            \begin{sec3lemma1}\label{lab_sec3_lemma1}
            Define a mapping $ f $, such that for $ k\in D_{0}^{(pq)}\cup D_{1}^{(pq)},\  k\mapsto f(k)=(k\bmod p,\ k\bmod q) $. Then, the following mappings deduced from the mapping $ f $ are bijective.
            \begin{enumerate} 
            \item
            \begin{equation*}
            f:D_{0}^{(pq)}\rightarrow D_{0}^{(p)}\times D_{0}^{(q)}\cup D_{1}^{(p)}\times D_{1}^{(q)},
            \end{equation*}
            \item
            \begin{equation*}
            f:D_{1}^{(pq)}\rightarrow D_{0}^{(p)}\times D_{1}^{(q)}\cup D_{1}^{(p)}\times D_{0}^{(q)}.
            \end{equation*}
            \end{enumerate}
            \end{sec3lemma1}
            \begin{proof}
            We only prove Lemma \ref{lab_sec3_lemma1} (1) since the proof for Lemma \ref{lab_sec3_lemma1} (2) is similar. At first, we prove the range of $ f $ is such that $ \text{Ran}(f) \subseteq D_{0}^{(p)}\times D_{0}^{(q)}\cup D_{1}^{(p)}\times D_{1}^{(q)}$. Let $ k_{1}= k\bmod p$, and $k_{2}= k\bmod q$. Without loss of generality, suppose $ (k_{1},\ k_{2}) \in D_{0}^{(p)}\times D_{1}^{(q)}$, and we go to prove a contradiction. Since $ k\in D_{0}^{(pq)} $, there is a $ n $ such that $ k\equiv g^{n}\pmod{pq} $, where $ 0\leq n<\frac{(p-1)(q-1)}{2} $. There must be a pair of integers $ i $ and $ j $, with $ 0\leq i<\frac{(p-1)}{2}  $ and $ 0\leq j<\frac{(q-1)}{2}  $, such that $ k_{1}\equiv g^{2i}\pmod{p} $, and $ k_{2}\equiv g^{2j+1}\pmod{q} $. From above discussion, we have the following congruent system:
            \begin{equation}\label{sec3_lemma1_proof_eq1}
            \begin{cases}
            g^{n}\equiv g^{2i}  &\pmod{p},\\
            g^{n}\equiv g^{2j+1}&\pmod{q}.
            \end{cases}
            \end{equation}
            From (\ref{sec3_lemma1_proof_eq1}), it follows 
            \begin{equation}\label{sec3_lemma1_proof_eq2}
            \begin{cases}
            n\equiv 2i  &\pmod{p-1},\\
            n\equiv 2j+1  &\pmod{q-1}.
            \end{cases}
            \end{equation}
            Remark that (\ref{sec3_lemma1_proof_eq2}) implies that $ gcd(p-1,q-1) =1 $, which is contradictory to the predefined condition $ gcd(p-1,q-1) =2$.
            
            By the Chinese Remainder Theorem, it is obvious that the mapping $ f $ is injective. We go to prove that $ f $ is surjective as well. Recall the meaning of the symbols used in the above proof. Let $ s(q-1)+t(p-1)=2 $, and $ (k_{1},\ k_{2}) \in D_{0}^{(p)}\times D_{0}^{(q)} $. Hence, $ k_{1}\equiv g^{2i}\pmod{p} $,  and $k_{2}\equiv g^{2j}\pmod{q}$. Resolve the following congruent system:
            \begin{equation}\label{sec3_lemma1_proof_eq3}
            \begin{cases}
            n\equiv 2i  &\pmod{p-1},\\
            n\equiv 2j  &\pmod{q-1}.
            \end{cases}
            \end{equation}
            By the Generalized Chinese Remainder Theorem, the solution of (\ref{sec3_lemma1_proof_eq3}) is equal to
            \begin{equation}\label{sec3_lemma1_proof_eq4}
            n\equiv s(q-1)i+t(p-1)j \pmod{\dfrac{(p-1)(q-1)}{2}}.
            \end{equation}
            It is clear that $ g^{n}\pmod{pq}\in D_{0}^{(pq)} $.
            \end{proof}
            
        \begin{sec3lemma2}\label{lab_sec3_lemma2}
          Let $ f $ be a mapping , such that for $ k\in D_{0}^{(2pq)}\cup D_{1}^{(2pq)},\  k\mapsto f(k)=(k\pmod 2,\ k\pmod{pq} )= (1,\ k\pmod{pq} )$. Then, the following mappings related to $ f $ are bijective.
          \begin{enumerate} 
          \item
          \begin{equation*}
          f:D_{0}^{(2pq)}\rightarrow Z^{*}_{2}\times D_{0}^{(pq)}=\lbrace 1\rbrace \times D_{0}^{(pq)},
          \end{equation*}
          \item
          \begin{equation*}
          f:D_{1}^{(2pq)}\rightarrow Z^{*}_{2}\times D_{1}^{(pq)}=\lbrace 1\rbrace \times D_{1}^{(pq)}.
          \end{equation*}
          \end{enumerate}
          \end{sec3lemma2}
          \begin{proof}
          The proof is simple and similar to that of Lemma \ref{lab_sec3_lemma1}.
          \end{proof}
          \begin{sec3lemma3}\label{lab_sec3_lemma3}
          Let $ f $ be a mapping , such that for $ k\in D_{0}^{(2p)}\cup D_{1}^{(2p)},\  k\mapsto f(k)=(k\pmod 2,\ k\pmod{p} )= (1,\ k\pmod{p} )$. Then, the following mappings related to $ f $ are bijective.
          \begin{enumerate} 
          \item
          \begin{equation*}
          f:D_{0}^{(2p)}\rightarrow Z^{*}_{2}\times D_{0}^{(p)}=\lbrace 1\rbrace \times D_{0}^{(p)},
          \end{equation*}
          \item
          \begin{equation*}
          f:D_{1}^{(2p)}\rightarrow Z^{*}_{2}\times D_{1}^{(p)}=\lbrace 1\rbrace \times D_{1}^{(p)}.
          \end{equation*}
          \end{enumerate}
          \end{sec3lemma3}
          \begin{proof}
          It is obvious.
          \end{proof}
          \begin{sec3lemma4}\label{lab_sec3_lemma4}
          \begin{equation}\label{sec3_lemma4_proof_eq1}
          2_{T}^{-1}\in
          \begin{cases}
          D_{0}^{(T)}&(2,2)\in D_{0}^{(p)}\times D_{0}^{(q)}\cup D_{1}^{(p)}\times D_{1}^{(q)},\\
          D_{1}^{(T)}&(2,2)\in D_{0}^{(p)}\times D_{1}^{(q)}\cup D_{1}^{(p)}\times D_{0}^{(q)}.
          \end{cases}
          \end{equation}
          Recall that $ T=pq $, and $ 2_{T}^{-1}=2^{-1}\bmod T$.
          \end{sec3lemma4}
          \begin{proof}
          It is clear that $ 2_{T}^{-1}\equiv 2_{p}^{-1} \bmod p$, and $ 2_{T}^{-1}\equiv 2_{q}^{-1} \bmod q$. The conclusion follows from Lemma \ref{lab_sec2_lemma3x} and Lemma \ref{lab_sec3_lemma1}.
          \end{proof}
          
           We define six integer functions in order to simplify notation.
              \begin{equation}\label{sec3_eq_parameters}
              \begin{cases}
              &A_{0}(k)= \sum_{i\in D_{0}^{(p)}}\beta^{ki}_{p},
              \ A_{1}(k)= \sum_{i\in D_{1}^{(p)}}\beta^{ki}_{p};\\
              &B_{0}(k)= \sum_{i\in D_{0}^{(q)}}\beta^{ki}_{q},
              \ B_{1}(k)= \sum_{i\in D_{1}^{(q)}}\beta^{ki}_{q},\\
              &Z^{(p)}(k)=A_{0}(k)+A_{1}(k),\\
              &Z^{(q)}(k)=B_{0}(k)+B_{1}(k).
              \end{cases}
              \end{equation}
              Where $ k\in Z_{2T} $.
              \begin{sec3lemma5}\label{lab_sec3_lemma5}
              \begin{enumerate} 
              \item
              \begin{equation*}
              \begin{split}
              \sum_{n\in D_{0}^{(2pq)}}\beta^{kn}=\beta_{2}^{k}\bigl(A_{0}(2_{p}^{-1}q_{p}^{-1}k)B_{0}(2_{q}^{-1}p_{q}^{-1}k)+
              &A_{1}(2_{p}^{-1}q_{p}^{-1}k)B_{1}(2_{q}^{-1}p_{q}^{-1}k)\bigr).
              \end{split}
              \end{equation*}
              \item
              \begin{equation*}
              \begin{split}
              \sum_{n\in D_{1}^{(2pq)}}\beta^{kn}=\beta_{2}^{k}\bigl(A_{0}(2_{p}^{-1}q_{p}^{-1}k)B_{1}(2_{q}^{-1}p_{q}^{-1}k)+
              &A_{1}(2_{p}^{-1}q_{p}^{-1}k)B_{0}(2_{q}^{-1}p_{q}^{-1}k)\bigr).
              \end{split}
              \end{equation*}
              \end{enumerate}
              Where $ k\in Z_{2T} $.
              \end{sec3lemma5}
              \begin{proof}
              We give only the proof for case (1). By Lemma \ref{lab_sec3_lemma2} and Lemma \ref{lab_sec3_lemma1}, $ n\equiv pq+2\cdot 2_{T}^{-1}\bigl(M_{p}k_{1}+M_{q}k_{2}\bigr) \pmod{2T}$, with $ n\in D_{0}^{(2pq)} $, $ k_{1}=n\bmod p $, $ k_{2}=n\bmod q $, and $(k_{1},\ k_{2})\in D_{0}^{(p)}\times D_{0}^{(q)}\cup D_{1}^{(p)}\times D_{1}^{(q)} $.
              \begin{equation*}
              \begin{split}
              \sum_{n\in D_{0}^{(2pq)}}\beta^{kn}&=\beta_{2}^{k}\bigl(\sum_{(k_{1},k_{2})\in D_{0}^{(p)}\times D_{0}^{(q)}}\beta_{T}^{k2_{T}^{-1}\bigl(M_{p}k_{1}+M_{q}k_{2}\bigr)}+
              \sum_{(k_{1},k_{2})\in D_{1}^{(p)}\times D_{1}^{(q)}}\beta_{T}^{k2_{T}^{-1}\bigl(M_{p}k_{1}+M_{q}k_{2}\bigr)}
              \bigr)\\
              &=\beta_{2}^{k}\bigl(
              \sum_{k_{1}\in D_{0}^{(p)}}\beta_{p}^{2^{-1}_{p}q^{-1}_{p}kk_{1}}\sum_{k_{2}\in D_{0}^{(q)}}\beta_{q}^{2^{-1}_{q}p^{-1}_{q}kk_{2}}+
              \sum_{k_{1}\in D_{1}^{(p)}}\beta_{p}^{2^{-1}_{p}q^{-1}_{p}kk_{1}}\sum_{k_{2}\in D_{1}^{(q)}}\beta_{q}^{2^{-1}_{q}p^{-1}_{q}kk_{2}}
              \bigr).
              \end{split}
              \end{equation*}
              \end{proof}
            \begin{sec3lemma6}\label{lab_sec3_lemma6}
            \begin{enumerate} 
            \item
            \begin{equation*}
            \begin{split}
              &\sum_{n\in qD_{0}^{(2p)}}\beta^{kn}=\beta_{2}^{k}A_{0}(2^{-1}_{p}k),\\
              &\sum_{n\in qD_{1}^{(2p)}}\beta^{kn}=\beta_{2}^{k}A_{1}(2^{-1}_{p}k).
             \end{split}
            \end{equation*}
            \item
            \begin{equation*}
             \begin{split}
            &\sum_{n\in pD_{0}^{(2q)}}\beta^{kn}=\beta_{2}^{k}B_{0}(2^{-1}_{q}k),\\
            &\sum_{n\in pD_{1}^{(2q)}}\beta^{kn}=\beta_{2}^{k}B_{1}(2^{-1}_{q}k).
            \end{split}
            \end{equation*}
            \end{enumerate}
            Where $ k\in Z_{2T} $.
            \end{sec3lemma6}
            \begin{proof}
            Using Lemma \ref{lab_sec3_lemma3} both for $ p $ and $ q $, and the same proof method as that for Lemma \ref{lab_sec3_lemma5}, we can obtain the conclusion of the actual lemma.
            \end{proof}
            \begin{sec3lemma7}\label{lab_sec3_lemma7}
            \begin{enumerate}
            \item
            If $ \bigl(\tfrac{q}{p}\bigr)=\bigl(\tfrac{p}{q}\bigr) $, then
            \begin{equation*}
            \begin{split}
            \sum_{i\in D_{0}^{(T)}}\beta_{T}^{ki}&= A_{0}(k)B_{0}(k)+A_{1}(k)B_{1}(k),\\
            \sum_{i\in D_{1}^{(T)}}\beta_{T}^{ki}&= A_{0}(k)B_{1}(k)+A_{1}(k)B_{0}(k).
            \end{split}
            \end{equation*}
            \item
             If $ \bigl(\tfrac{q}{p}\bigr)=-\bigl(\tfrac{p}{q}\bigr) $, then
             \begin{equation*}
             \begin{split}
              \sum_{i\in D_{0}^{(T)}}\beta_{T}^{ki}&= A_{0}(k)B_{1}(k)+A_{1}(k)B_{0}(k),\\
              \sum_{i\in D_{1}^{(T)}}\beta_{T}^{ki}&= A_{0}(k)B_{0}(k)+A_{1}(k)B_{1}(k).
              \end{split}
             \end{equation*}
            \end{enumerate}
            \end{sec3lemma7}
            \begin{proof}
            We only prove the expansion formula for $ \sum_{i\in D_{1}^{(T)}}\beta_{T}^{ki} $ in case (1).  Let $ i\equiv M_{p}m+ M_{q}n \pmod{pq}$ where $ m\equiv i\pmod p $, $ n\equiv i\pmod q $, and $ i\in D_{1}^{(T)} $. We distinguish two cases $\bigl(\tfrac{q}{p}\bigr)=\bigl(\tfrac{p}{q}\bigr)=1  $ and $\bigl(\tfrac{q}{p}\bigr)=\bigl(\tfrac{p}{q}\bigr)=-1  $. It is clear that $ q_{p}^{-1}=q^{-1} \pmod p \in D_{0}^{(p)} $ if $\bigl(\tfrac{q}{p}\bigr)=1$, and $ q_{p}^{-1} \in D_{1}^{(p)} $ if $\bigl(\tfrac{q}{p}\bigr)=-1$, according to the Legendre Symbol definition. Same situation for $ p_{q}^{-1} $. We use Lemma \ref{lab_sec3_lemma1} to decompose the summation over $ D_{1}^{(T)} $ into the summations over $ D_{0}^{(p)}\times D_{1}^{(q)}\cup D_{1}^{(p)}\times D_{0}^{(q)} $.
            \begin{enumerate}
            \item $\bigl(\tfrac{q}{p}\bigr)=\bigl(\tfrac{p}{q}\bigr)=1  $.
            \begin{equation*}
            \begin{split}
             \sum_{i\in D_{1}^{(T)}}\beta_{T}^{ki}&=\sum_{m\in D_{0}^{(p)} }\beta_{T}^{kM_{p}m}\cdot\sum_{n\in D_{1}^{(q)} }\beta_{T}^{kM_{q}n}+
             \sum_{m\in D_{1}^{(p)} }\beta_{T}^{kM_{p}m}\cdot\sum_{n\in D_{0}^{(q)} }\beta_{T}^{kM_{q}n}\\
             &=\sum_{m\in D_{0}^{(p)} }(\beta_{T}^{q})^{k q_{p}^{-1}m}\cdot\sum_{n\in D_{1}^{(q)} }(\beta_{T}^{p})^{kp_{q}^{-1}n}+
             \sum_{m\in D_{1}^{(p)} }(\beta_{T}^{q})^{k q_{p}^{-1}m}\cdot\sum_{n\in D_{0}^{(q)} }(\beta_{T}^{p})^{kp_{q}^{-1}n}\\
            \end{split}
            \end{equation*}
            Note that $ \beta_{T}^{q}= \beta_{p} $ and $ \beta_{T}^{p}= \beta_{q} $. Using Lemma \ref{lab_sec2_lemma2x}, we eliminate the exponents $ p_{q}^{-1} $ and $ q_{p}^{-1} $ from  above formula:
            \begin{equation*}
            \begin{split}
            \sum_{i\in D_{1}^{(T)}}\beta_{T}^{ki}&=\sum_{m\in q_{p}^{-1}D_{0}^{(p)} }(\beta_{p})^{km}\cdot\sum_{n\in p_{q}^{-1}D_{1}^{(q)} }(\beta_{q})^{kn}+
              \sum_{m\in q_{p}^{-1}D_{1}^{(p)} }(\beta_{p})^{km}\cdot\sum_{n\in p_{q}^{-1}D_{0}^{(q)} }(\beta_{q})^{kn}\\
            &=\sum_{m\in D_{0}^{(p)} }(\beta_{p})^{km}\cdot\sum_{n\in D_{1}^{(q)} }(\beta_{q})^{kn}+
              \sum_{m\in D_{1}^{(p)} }(\beta_{p})^{km}\cdot\sum_{n\in D_{0}^{(q)} }(\beta_{q})^{kn}\\
              &=A_{0}(k)B_{1}(k)+A_{1}(k)B_{0}(k).
            \end{split}
            \end{equation*} 
            \item  $\bigl(\tfrac{q}{p}\bigr)=\bigl(\tfrac{p}{q}\bigr)=-1  $. Note that in this case, $ q_{p}^{-1} \in D_{1}^{(p)} $  and  $ p_{q}^{-1} \in D_{1}^{(q)} $. The rest of the proof is same as for $\bigl(\tfrac{q}{p}\bigr)=\bigl(\tfrac{p}{q}\bigr)=1  $.
            
            \end{enumerate}
            
            \end{proof}

            By the definition of generating polynomial in (\ref{lab_eq_GCD_GP_ex}), we can  write down the one of the sequence defined in (\ref{lab_seq_chang_ex}):
            \begin{equation}\label{lab_seq_chang_generating_polynomial}
            \begin{split}
            G_{s}(t)=&2t^{T}+\bigl(\sum_{i\in D_{1}^{(2T)}}+\sum_{i\in qD_{1}^{(2p)}}+\sum_{i\in pD_{1}^{(2q)}}\bigr)t^{i}+\\
            &2\bigl(\sum_{i\in 2D_{0}^{(T)}}+\sum_{i\in 2qD_{0}^{(p)}}+\sum_{i\in 2pD_{0}^{(q)}}\bigr)t^{i}+\\
            &3\bigl(\sum_{i\in 2D_{1}^{(T)}}+\sum_{i\in 2qD_{1}^{(p)}}+\sum_{i\in 2pD_{1}^{(q)}}\bigr)t^{i}.
            \end{split}
            \end{equation}
            
            Since $ \beta $ is the $ 2T^{th} $ root of unity in the extension field $ \mathbf{GF}(r^{m}) $, all the zeros of the polynomial $ x^{2T}-1 $ over $ \mathbf{GF}(r^{m})[x]  $ are under the form $ \beta^{k} $, with $ k\in Z_{2T} $. If the same $ \beta^{k} $ is also the zero of the generating polynomial in (\ref{lab_seq_chang_generating_polynomial}), it simply means that $ x- \beta^{k} $ is a common factor of the $ G_{s}(x) $ and $ x^{2T}-1 $ . By explicitly compute out all the zeros of the generating polynomial $ G_{s}(x) $ under the form $ \beta^{k} $, we can determine the degree of the corresponding minimal polynomial, from which the linear complexity of the sequence in (\ref{lab_seq_chang_ex}) is straightforward.
            
            From (\ref{lab_seq_chang_generating_polynomial}) and by  applying Lemma \ref{lab_sec2_lemma2x}, Lemma \ref{lab_sec3_lemma5}-\ref{lab_sec3_lemma6} , we can get the explicit algebraic form of $ G_{s}(\beta^{k}) $:
            \begin{equation}\label{lab_seq_chang_GP_in_beta_k}
            \begin{split}
             G_{s}(\beta^{k})=2\beta_{2}^{k}&+\biggl(\beta_{2}^{k}\sum_{i\in D_{1}^{(T)}}\beta_{T}^{2_{T}^{-1}ki}+\sum_{i\in D_{1}^{(T)}}\beta_{T}^{ki}\biggr)\\
             &+\biggl(\beta_{2}^{k}\sum_{i\in D_{1}^{(p)}}\beta_{p}^{2_{p}^{-1}ki}+\sum_{i\in D_{1}^{(p)}}\beta_{p}^{ki}\biggr)\\
             &+\biggl(\beta_{2}^{k}\sum_{i\in D_{1}^{(q)}}\beta_{q}^{2_{q}^{-1}ki}+\sum_{i\in D_{1}^{(q)}}\beta_{q}^{ki}\biggr)\\
             &+2\biggl(Z^{(p)}(k)\cdot Z^{(q)}(k)+Z^{(p)}(k)+Z^{(q)}(k)\biggr).
            \end{split}
            \end{equation}
            Where $ k\in Z_{2T} $.
            
   \subsection{$ 2 $ is a quadratic residue both modulo $ p $ and modulo $ q $}  
    
     Let $ k_{1}=k\bmod p $ and $ k_{2}=k\bmod q $. Next lemma gives the values of $ G_{s}(\beta^{k}) $ for $ k\in Z_{2T} $.
     \begin{sec3lemma8}\label{lab_sec3_lemma8}
     If $ (2,\ 2)\in D_{0}^{(p)}\times D_{0}^{(q)} $, then the values of $ G_{s}(\beta^{k}) $ where $ k\in Z_{2T} $ can be determined as follows:
     \begin{enumerate} 
     \item $ k\in \lbrace 0,\ pq \rbrace \cup Z^{*}_{2T}\cup qZ^{*}_{2p}\cup pZ^{*}_{2q} $.
     \begin{equation*}
     G_{s}(\beta^{k})=
     \begin{cases}
     3pq-1&k\in\lbrace 0\rbrace,\\
     2pq-4&k\in\lbrace pq\rbrace,\\
     -4   &k\in Z^{*}_{2T}\cup qZ^{*}_{2p}\cup pZ^{*}_{2q}.
     \end{cases}
     \end{equation*}
     \item $ k\in 2D_{0}^{(T)}\cup 2D_{1}^{(T)} $.
     \begin{enumerate}
     \item 
     If $ k\in 2D^{(T)}_{0} $ and $ (k_{1},\ k_{2})\in D^{(p)}_{0}\times D^{(q)}_{0} $ and $ \bigl(\tfrac{p}{q}\bigr)=-\bigl(\tfrac{q}{p}\bigr) $, then
     \begin{equation*}
     G_{s}(\beta^{k})=2(A_{0}B_{0}+A_{1}B_{1}+A_{1}+B_{1}).
     \end{equation*}
     \item 
     If $ k\in 2D^{(T)}_{0} $ and $ (k_{1},\ k_{2})\in D^{(p)}_{0}\times D^{(q)}_{0} $ and $ \bigl(\tfrac{p}{q}\bigr)=\bigl(\tfrac{q}{p}\bigr) $, then
     \begin{equation*}
     G_{s}(\beta^{k})=2(A_{0}B_{1}+A_{1}B_{0}+A_{1}+B_{1}).
     \end{equation*}
     \item 
     If $ k\in 2D^{(T)}_{0} $ and $ (k_{1},\ k_{2})\in D^{(p)}_{1}\times D^{(q)}_{1} $ and $ \bigl(\tfrac{p}{q}\bigr)=-\bigl(\tfrac{q}{p}\bigr) $, then
     \begin{equation*}
     G_{s}(\beta^{k})=2(A_{0}B_{0}+A_{1}B_{1}+A_{0}+B_{0}).
     \end{equation*}
     \item 
     If $ k\in 2D^{(T)}_{0} $ and $ (k_{1},\ k_{2})\in D^{(p)}_{1}\times D^{(q)}_{1} $ and $ \bigl(\tfrac{p}{q}\bigr)=\bigl(\tfrac{q}{p}\bigr) $, then
     \begin{equation*}
     G_{s}(\beta^{k})=2(A_{0}B_{1}+A_{1}B_{0}+A_{0}+B_{0}).
     \end{equation*}
     \item 
     If $ k\in 2D^{(T)}_{1} $ and $ (k_{1},\ k_{2})\in D^{(p)}_{0}\times D^{(q)}_{1} $ and $ \bigl(\tfrac{p}{q}\bigr)=-\bigl(\tfrac{q}{p}\bigr) $, then
     \begin{equation*}
     G_{s}(\beta^{k})=2(A_{0}B_{1}+A_{1}B_{0}+A_{1}+B_{0}).
     \end{equation*}
     \item 
     If $ k\in 2D^{(T)}_{1} $ and $ (k_{1},\ k_{2})\in D^{(p)}_{0}\times D^{(q)}_{1} $ and $ \bigl(\tfrac{p}{q}\bigr)=\bigl(\tfrac{q}{p}\bigr) $, then
     \begin{equation*}
     G_{s}(\beta^{k})=2(A_{0}B_{0}+A_{1}B_{1}+A_{1}+B_{0}).
     \end{equation*}
     \item 
     If $ k\in 2D^{(T)}_{1} $ and $ (k_{1},\ k_{2})\in D^{(p)}_{1}\times D^{(q)}_{0} $ and $ \bigl(\tfrac{p}{q}\bigr)=-\bigl(\tfrac{q}{p}\bigr) $, then
     \begin{equation*}
     G_{s}(\beta^{k})=2(A_{0}B_{1}+A_{1}B_{0}+A_{0}+B_{1}).
     \end{equation*}
     \item 
     If $ k\in 2D^{(T)}_{1} $ and $ (k_{1},\ k_{2})\in D^{(p)}_{1}\times D^{(q)}_{0} $ and $ \bigl(\tfrac{p}{q}\bigr)=\bigl(\tfrac{q}{p}\bigr) $, then
     \begin{equation*}
     G_{s}(\beta^{k})=2(A_{0}B_{0}+A_{1}B_{1}+A_{0}+B_{1}).
     \end{equation*}
     \end{enumerate}
     \item $ k\in 2qD_{0}^{(p)}\cup 2qD_{1}^{(p)} $.
     \begin{enumerate}
     \item
     If $ k\in 2qD_{0}^{(p)} $ and $ \bigl(\tfrac{q}{p}\bigr)=-1 $ or $ k\in 2qD_{1}^{(p)} $ and $ \bigl(\tfrac{q}{p}\bigr)=1 $, then
     \begin{equation*}
     G_{s}(\beta^{k})=2A_{0}.
     \end{equation*}
     \item
     If $ k\in 2qD_{0}^{(p)} $ and $ \bigl(\tfrac{q}{p}\bigr)=1 $ or $ k\in 2qD_{1}^{(p)} $ and $ \bigl(\tfrac{q}{p}\bigr)=-1 $, then
     \begin{equation*}
     G_{s}(\beta^{k})=2A_{1}.
     \end{equation*}
     \end{enumerate}
     \item $ k\in 2pD_{0}^{(q)}\cup 2pD_{1}^{(q)} $.
     \begin{enumerate}
     \item
     If $ k\in 2pD_{0}^{(q)} $ and $ \bigl(\tfrac{p}{q}\bigr)=-1 $ or $ k\in 2pD_{1}^{(q)} $ and $ \bigl(\tfrac{p}{q}\bigr)=1 $, then
     \begin{equation*}
     G_{s}(\beta^{k})=2B_{0}.
     \end{equation*}
     \item
     If $ k\in 2pD_{0}^{(q)} $ and $ \bigl(\tfrac{p}{q}\bigr)=1 $ or $ k\in 2pD_{1}^{(q)} $ and $ \bigl(\tfrac{p}{q}\bigr)=-1 $, then
     \begin{equation*}
     G_{s}(\beta^{k})=2B_{1}.
     \end{equation*}
     \end{enumerate}
     \end{enumerate}
     \end{sec3lemma8}
     \begin{proof}
     Using Lemma \ref{lab_sec3_lemma4} and Lemma \ref{lab_sec2_lemma2x} to eliminate the exponents $ 2_{T}^{-1} $,\  $ 2_{p}^{-1} $, and $ 2_{q}^{-1} $ from (\ref{lab_seq_chang_GP_in_beta_k}), we get a new form of (\ref{lab_seq_chang_GP_in_beta_k}):
     \begin{equation}\label{lab_seq_in_beta_k_proof}
     \begin{split}
      &G_{s}(\beta^{k})=2\beta_{2}^{k}+\\
      &\biggl(\beta_{2}^{k}+1\biggr)\biggl(\sum_{i\in D_{1}^{(T)}}\beta_{T}^{ki}+\sum_{i\in D_{1}^{(p)}}\beta_{p}^{ki}+\sum_{i\in D_{1}^{(q)}}\beta_{q}^{ki}\biggr)\\
      &+2\biggl(Z^{(p)}(k)\cdot Z^{(q)}(k)+Z^{(p)}(k)+Z^{(q)}(k)\biggr).
     \end{split}
     \end{equation}
     Where $ k\in Z_{2T} $.
     
     In (\ref{lab_seq_in_beta_k_proof}), let $ s_{1}=\sum_{i\in D_{1}^{(T)}}\beta_{T}^{ki}+\sum_{i\in D_{1}^{(p)}}\beta_{p}^{ki}+\sum_{i\in D_{1}^{(q)}}\beta_{q}^{ki} $, $s_{2}= Z^{(p)}(k) $, $s_{3}= Z^{(q)}(k) $, $s_{4}= Z^{(p)}(k)\cdot Z^{(q)}(k) $, and $ s_{5}= \beta_{2}^{k}$. It is easy to compute  the values of $ s_{2}$, and $ s_{3} $, for all $ k\in Z_{2T} $, see below:
     \begin{equation}\label{lab_sec3_L3_prf_s2}
     s_{2}= 
     \begin{cases}
     -1  &k\in Z_{2T}^{*}\cup 2Z_{T}^{*}\cup qZ_{2p}^{*}\cup 2qZ_{p}^{*},\\
     p-1 &k\in\lbrace 0,T\rbrace \cup pZ_{2q}^{*}\cup 2pZ_{q}^{*}.
     \end{cases}
     \end{equation}
     \begin{equation}\label{lab_sec3_L3_prf_s3}
     s_{3}= 
     \begin{cases}
     -1  &k\in Z_{2T}^{*}\cup 2Z_{T}^{*}\cup pZ_{2q}^{*}\cup 2pZ_{q}^{*},\\
     q-1 &k\in\lbrace 0,T\rbrace \cup qZ_{2p}^{*}\cup 2qZ_{p}^{*}.
     \end{cases}
     \end{equation}
     After having computed $ s_{2} $ and $ s_{3} $, it is straightforward to find the values of $ s_{4} $. On the other hand, computation of $ s_{5} $ is trivial. Due to limited space and similarity of proof process, we consider only the case (3) of Lemma \ref{lab_sec3_lemma8}, that is, $ k\in 2qZ_{p}^{*} $. 
     
     By Lemma \ref{lab_sec2_lemma2x} $ A_{1}(k) =A_{1}$ if $ k\in D_{0}^{(p)} $, and $ A_{1}(k) =A_{0}$ if $ k\in D_{1}^{(p)} $.
     
     Remark that if  $ \bigl(\tfrac{q}{p}\bigr)=1 $, then $ q \pmod p \in D_{0}^{(p)}$, else $ q \pmod p \in D_{1}^{(p)}$. Next, we compute $ s_{1} $ for  $ k\in 2qZ_{p}^{*} $. Let $ k=2qk_{1} $,  where $ k_{1}=k \pmod{p} $. It is obvious that $ B_{0}(k) = B_{1}(k)=\tfrac{q-1}{2}$, $ A_{0}(k)+A_{1}(k)=s_{2}=Z^{(p)}(k)=-1 $, $ B_{0}(k)+B_{1}(k)=s_{3}=Z^{(q)}(k)=q-1 $, and $ s_{5}=\beta_{2}^{k}=1 $. By Lemma \ref{lab_sec3_lemma7}, we distinguish two cases:
     \begin{enumerate}
     \item $ \bigl(\tfrac{q}{p}\bigr)=\bigl(\tfrac{p}{q}\bigr) $.
     \begin{equation*}
     \begin{split}
     s_{1}&=\sum_{i\in D_{1}^{(T)}}\beta_{T}^{ki}+\sum_{i\in D_{1}^{(p)}}\beta_{p}^{ki}+\sum_{i\in D_{1}^{(q)}}\beta_{q}^{ki} \\
     &=A_{0}(k)B_{1}(k)+A_{1}(k)B_{0}(k)+A_{1}(k)+B_{1}(k)\\
     &=(A_{0}(k)+A_{1}(k))\cdot\frac{q-1}{2}+A_{1}(k)+\frac{q-1}{2}\\
     &=(-1)\cdot\frac{q-1}{2}+A_{1}(k)+\frac{q-1}{2}=A_{1}(k).
     \end{split}
     \end{equation*}
     Note that $ s_{1}=A_{1}(k)=A_{1}(2qk_{1})=A_{1}(qk_{1}) $ since $ 2\in  D_{0}^{(p)} $. Using Lemma \ref{lab_sec2_lemma2x}, possible combination for $ k_{1} $ and $ q\pmod p\in \lbrace D_{0}^{(p)},D_{1}^{(p)}\rbrace $ gives the following values to $ s_{1} $:
     \begin{equation}\label{lab_sec3_L3_prf_s4x}
        s_{1}=
        \begin{cases}
        A_{0}&k_{1}\in D_{0}^{(p)}\,\text{and}\,\bigl(\frac{q}{p}\bigr)=-1\,\text{or}\,\\
        &k_{1}\in D_{1}^{(p)}\,\text{and}\,\bigl(\frac{q}{p}\bigr)=1,\\
        A_{1}&k_{1}\in D_{0}^{(p)}\,\text{and}\,\bigl(\frac{q}{p}\bigr)=1\,\text{or}\,\\
        &k_{1}\in D_{1}^{(p)}\,\text{and}\,\bigl(\frac{q}{p}\bigr)=-1.
        \end{cases}
        \end{equation}
     \item $ \bigl(\tfrac{q}{p}\bigr)=-\bigl(\tfrac{p}{q}\bigr) $.
      After rather similar computation, we find that the final values of $ s_{1} $ in this case are the same as that listed in (\ref{lab_sec3_L3_prf_s4x}).
     \end{enumerate}
     
     From (\ref{lab_seq_in_beta_k_proof})-(\ref{lab_sec3_L3_prf_s4x}), we obtain
     \begin{equation*}
     \begin{split}
     &G_{s}(\beta^{k})=2\beta_{2}^{k}+\\
         &\biggl(\beta_{2}^{k}+1\biggr)\biggl(\sum_{i\in D_{1}^{(T)}}\beta_{T}^{ki}+\sum_{i\in D_{1}^{(p)}}\beta_{p}^{ki}+\sum_{i\in D_{1}^{(q)}}\beta_{q}^{ki}\biggr)\\
         &+2\biggl(Z^{(p)}(k)\cdot Z^{(q)}(k)+Z^{(p)}(k)+Z^{(q)}(k)\biggr)\\
         &=2\cdot s_{5}+(1+s_{5})\cdot s_{1}+2(s_{2}\cdot s_{3}+s_{2}+s_{3})\\
         &=2\cdot 1+(1+1)\cdot s_{1}+2((-1)\cdot(q-1)+(-1)+(q-1))\\
         &=2 s_{1}.
     \end{split}
     \end{equation*}
     
     Hence, for $ k\in 2qZ_{p}^{*} $ we get 
     \begin{equation*}
     G_{s}(\beta^{k})=2 s_{1}=
     \begin{cases}
           2A_{0}&k_{1}\in D_{0}^{(p)}\,\text{and}\,\bigl(\frac{q}{p}\bigr)=-1\,\text{or}\,\\
           &k_{1}\in D_{1}^{(p)}\,\text{and}\,\bigl(\frac{q}{p}\bigr)=1,\\
           2A_{1}&k_{1}\in D_{0}^{(p)}\,\text{and}\,\bigl(\frac{q}{p}\bigr)=1\,\text{or}\,\\
           &k_{1}\in D_{1}^{(p)}\,\text{and}\,\bigl(\frac{q}{p}\bigr)=-1,
           \end{cases}
     \end{equation*}
     that correspond to the case (3) in Lemma \ref{lab_sec3_lemma8}.
     \end{proof} 
     
     We below list  main theorems of this subsection:
     \begin{sec3thm1}\label{lab_MainThorem_01}
     Let $ (2,2) \in  D_{0}^{(p)}\times D_{0}^{(q)}$, $ \bigl(\tfrac{r}{p}\bigr) =1 $ and $ \bigl(\tfrac{r}{q}\bigr) =1 $. Then, the linear complexity over $ \mathbf{GF}(r^{m}) $ of the generalized cyclotomic quaternary sequences with period $ 2T $ specified in (\ref{lab_seq_chang_ex}), $  LC(s) $, can be determined according to next two cases.
     \begin{enumerate}
     \item $ (p,q)\in (1,-1) $ and $ r\mid p-1 $ and $ r\mid q+1 $ \textbf{OR} $ (p,q)\in (-1,1) $ and $ r\mid p+1 $ and $ r\mid q-1 $:
     \begin{equation*}
     LC(s)=\dfrac{5pq+p+q+1}{4}.
     \end{equation*}
     \item $ (p,q)\in (-1,-1) $ and $ r\mid p+1 $ and $ r\mid q+1 $:
     \begin{equation*}
     LC(s)=\dfrac{4pq-p-q+2}{2}.
     \end{equation*}
     \end{enumerate}
     \end{sec3thm1}
     \begin{proof}
       Since $ \bigl(\tfrac{r}{p}\bigr) =1 $ and $ \bigl(\tfrac{r}{q}\bigr) =1 $, it means that $ r\in  D_{0}^{(p)} $ and $ r\in  D_{0}^{(q)} $. Hence, by Lemma \ref{lab_sec2_lemma4x}, $ A_{0},A_{1},B_{0},B_{1} \in \mathbf{GF}(r)$. For Theorem \ref{lab_MainThorem_01} (1), we only prove the part where $ (p,\;q)\in (1,\;-1) $ and $ r\mid p-1 $ and $ r\mid q+1 $, since proof for another part is symmetric.
       
       Since $ (p,\;q)\in (1,\;-1) $, $ p\equiv 1 \pmod 4 $ and $ q\equiv 3 \pmod 4 $. Because $ r\mid p-1 $ and $ r\mid q+1 $, by Lemma \ref{lab_sec2_lemma1x}, $ A_{0},A_{1},B_{0},B_{1} \in \lbrace 0,\;-1\rbrace$.

       If $ A_{0} \in \lbrace 0,\;-1\rbrace $, by Lemma \ref{lab_sec3_lemma8} (3), $ G_{s}(\beta^{k})=0 $ for $ k\in 2qD_{0}^{(p)} $ or $ k\in 2qD_{1}^{(p)} $. It means that $ \prod_{k\in D}(x-\beta^{k}) $ are  common factors of both $ x^{2pq}-1 $ and $ G_{s}(x) $, where $ D=2qD_{0}^{(p)} $ or $ D=2qD_{1}^{(p)} $. The number of the common factors is equal to $ \frac{p-1}{2} $. On the other hand, if $ B_{0} \in \lbrace 0,\;-1\rbrace $, by Lemma \ref{lab_sec3_lemma8} (4), $ G_{s}(\beta^{k})=0 $ for $ k\in 2pD_{0}^{(q)} $ or $ k\in 2pD_{1}^{(q)} $. Hence, there are $ \frac{q-1}{2} $'s common factors between  $ x^{2pq}-1 $ and $ G_{s}(x) $, that are under the form $ \prod_{k\in D}(x-\beta^{k}) $ for $ D=2pD_{0}^{(q)} $ or $ D=2pD_{1}^{(q)} $.
       
       In order to find other zeros of $ G_{s}(\beta^{k}) $ with $ k\in Z_{2T} $, we define eight quantities that correspond to the eight sub-cases in Lemma \ref{lab_sec3_lemma8} (2):
       \begin{equation*}
       \begin{split}
       a&=2(A_{0}B_{0}+A_{1}B_{1}+A_{1}+B_{1}),\\
       b&=2(A_{0}B_{1}+A_{1}B_{0}+A_{1}+B_{1}),\\
       b&=2(A_{0}B_{0}+A_{1}B_{1}+A_{0}+B_{0}),\\
       d&=2(A_{0}B_{1}+A_{1}B_{0}+A_{0}+B_{0}),\\
       e&=2(A_{0}B_{1}+A_{1}B_{0}+A_{1}+B_{0}),\\
       f&=2(A_{0}B_{0}+A_{1}B_{1}+A_{1}+B_{0}),\\
       g&=2(A_{0}B_{1}+A_{1}B_{0}+A_{0}+B_{1}),\\
       h&=2(A_{0}B_{0}+A_{1}B_{1}+A_{0}+B_{1}).
       \end{split}
       \end{equation*}
       Note that for each $ (A_{0},\;B_{0}) \in \lbrace 0,\;-1\rbrace \times \lbrace 0,\;-1\rbrace$, from Table \ref{Lab_Table1}, there are three sub-cases in Lemma \ref{lab_sec3_lemma8} (2) where  $ G_{s}(\beta^{k})=0 $ for $ k\in 2D_{0}^{(T)}\cup 2D_{1}^{(T)} $, that contribute in total $ \frac{3(p-1)(q-1)}{4} $'s zeros to $ G_{s}(\beta^{k})$ for $ k\in Z_{2T}$. If we examine closely those three sub-cases, we could remark that it is required that $ \bigl(\tfrac{q}{p}\bigr) =\bigl(\tfrac{p}{q}\bigr) $ which is equivalent to $ (p,\;q)\in (1,\;-1)\cup (-1,\;1) $ according to the Quadratic Reciprocity Law (See Lemma \ref{lab_sec2_lemma3x} (2)).  
       
       From Lemma \ref{lab_sec3_lemma8} (1), it is obvious that $ G_{s}(\beta^{k})\ne 0 $ for $ k\in \lbrace 0,\ pq \rbrace \cup Z^{*}_{2T}\cup qZ^{*}_{2p}\cup pZ^{*}_{2q} $. By (\ref{lab_eq_GCD_LC_eq_ex}), 
       \begin{equation*}
       \begin{split}
       LC(s)&=2pq-\dfrac{3(p-1)(q-1)}{4}-\dfrac{p-1}{2}-\dfrac{q-1}{2}\\
            &=\dfrac{5pq+p+q+1}{4};
        \end{split}
       \end{equation*}
       and for Theorem \ref{lab_MainThorem_01} (2),
       \begin{equation*}
          \begin{split}
          LC(s)&=2pq-\dfrac{p-1}{2}-\dfrac{q-1}{2}\\
               &=\dfrac{4pq-p-q+2}{2}.
           \end{split}
          \end{equation*}
    \end{proof}
      \begin{table}[!t]
       \begin{threeparttable}[t]
       \renewcommand{\arraystretch}{1.3}
       \caption{Values of $ G_{s}(\beta^{k})$ in Lemma \ref{lab_sec3_lemma8}(2)\tnote{1}
       \label{Lab_Table1}}
       \centering
       \begin{tabular}{c|c|c|c|c|c|c|c|c|c|c|c}
       \hline
      $ A_{0} $ & $ A_{1} $ & $ B_{0} $ & $ B_{1} $ & $ a $ & $ b $ & $ c $ & $ d $ & $ e $ & $ f $ & $ g $ & $ h $\\
       \hline
       0 & -1 & 0 & -1 & -2 & -4 & 2 & 0 & -2 & 0 & -2 & 0\\
       \hline
       0 & -1 & -1 & 0 & -2 &  0 & -2 & 0 & -2 & -4 & 2 & 0\\
       \hline
       -1 & 0 & 0 & -1 & -2 &  0 & -2 & 0 & 2 & 0 & -2 & -4\\
       \hline
       -1 & 0 & -1 & 0 & 2 &  0 & -2 & -4 & -2 & 0 & -2 & 0\\
       \hline
       \end{tabular}
        \begin{tablenotes}
                   \item [1] $ \bigl(\tfrac{r}{p}\bigr) =1 $ and $ \bigl(\tfrac{r}{q}\bigr) =1 $.
        \end{tablenotes}
       \end{threeparttable}
       \end{table}
    \begin{algorithm}
      \caption{Algorithm to compute linear complexity}\label{lab_alg01}
      \begin{algorithmic}[1]
      \Procedure{LC}{$p,q,r$}
      \State Construct the set $D_{1}=D_{1}^{(2T)}\cup qD_{1}^{(2p)}\cup pD_{1}^{(2q)}$.
      \State Construct the set $D_{2}=2D_{0}^{(T)}\cup 2qD_{0}^{(p)}\cup 2pD_{0}^{(q)}$.
      \State Construct the set $D_{3}=2D_{1}^{(T)}\cup 2qD_{1}^{(p)}\cup 2pD_{1}^{(q)}$.
      \State Construct $ G_{s}(x) $ by (\ref{lab_seq_chang_generating_polynomial}) .
      \State Construct the polynomial $ P(x)=x^{2T}-1$.
      \State Compute $ \gcd(G_{s}(x), P(x))$.
      \State \textbf{return} $ 2pq-deg(\gcd(G_{s}(x), P(x))) $.
      \EndProcedure
      \end{algorithmic}
      \end{algorithm}
    \begin{sec3exp1}\label{lab_sec3exp1}
      We implement Algorithm \ref{lab_alg01} using the computer algebra system Maple, in particular, using the Maple built-in function \textbf{Gcdex} to compute the gcd of $ P(x) = x^{2T}-1$ and the generating polynomial $ G_{s}(x) $ defined in (\ref{lab_seq_chang_generating_polynomial}). Given a pair of primes $ p $ and $ q $, and $ r $ such that the conditions in Theorem \ref{lab_MainThorem_01} are satisfied, let $ LC_{Algo1} $ denote the linear complexity computed by Algorithm \ref{lab_alg01} and $ LC_{Th1} $ denote the linear complexity calculated using the formula in Theorem \ref{lab_MainThorem_01}. In Table \ref{Lab_Table2}, some numerical examples are listed. We observe that the numerical result by running the procedure in Algorithm \ref{lab_alg01} coincide the ones predicted by Theorem \ref{lab_MainThorem_01}.
      \end{sec3exp1}
      \begin{table}[!t]
         \begin{threeparttable}[t]
         \renewcommand{\arraystretch}{1.3}
         \caption{Linear Complexity Calculated By Theorem \ref{lab_MainThorem_01}}
         \label{Lab_Table2}
         \centering
         \begin{tabular}{c|c|c|c|c|c}
         \hline
        $ p $ & $ q $ & $ r $ & $ LC_{Algo1} $ & $ LC_{Th1} $ & \textbf{Condition}\\
         \hline
         41 & 79 & 5 & 4079 & 4079 & $ C_{1} $\tnote{a} \\
         \hline
         113 & 167 & 7 & 23659 & 23659 & $ C_{1} $ \\
         \hline
         89 & 263 & 11 & 29347 & 29347 & $ C_{1} $ \\
         \hline
         79 & 41 & 5 & 4079 & 4079 & $ C_{2} $\tnote{b} \\
         \hline
         167 & 113 & 7 & 23659 & 23659 & $ C_{2} $ \\
         \hline
         263 & 89 & 11 & 29347 & 29347 & $ C_{2} $ \\
         \hline
         311 & 313 & 13 & 121835 & 121835 & $ C_{2} $ \\
         \hline
         79 & 239 & 5 & 37604 & 37604 & $ C_{3} $\tnote{c} \\
         \hline
         167 & 223 & 7 & 74288 & 74288 & $ C_{3} $ \\
         \hline
         103 & 311 & 13 & 63860 & 63860 & $ C_{3} $ \\
         \hline
         \end{tabular}
         \begin{tablenotes}
         \item [a] $ (p,q)\in (1,-1) $ and $ r\mid p-1 $ and $ r\mid q+1 $.
         \item [b] $ (p,q)\in (-1,1) $ and $ r\mid p+1 $ and $ r\mid q-1 $.
         \item [c] $ (p,q)\in (-1,-1) $ and $ r\mid p+1 $ and $ r\mid q+1 $.
         \end{tablenotes}
         \end{threeparttable}
         \end{table}
   \begin{sec3thm1x}\label{lab_MainThorem_01x}
     Let $ (2,2) \in  D_{0}^{(p)}\times D_{0}^{(q)}$, $ \bigl(\tfrac{r}{p}\bigr) =1 $ and $ \bigl(\tfrac{r}{q}\bigr) =-1 $. Then, the linear complexity over $ \mathbf{GF}(r^{m}) $ of the generalized cyclotomic quaternary sequences with period $ 2T $ specified in (\ref{lab_seq_chang_ex}), $  LC(s) $, can be determined according to following cases.
      \begin{enumerate} 
      \item If $ (p,q)\in (1,-1) $ and $ r\mid p-1 $ and $ r\nmid 3q-1 $ and $ r\nmid 2q-4 $ \textbf{OR} $ (p,q)\in (-1,1) $ and $ r\mid p+1 $ and $ r\nmid 3q+1 $ and $ r\nmid 2q+4 $, then
      \begin{equation*}
         LC(s)=\frac{3pq+q}{2}.
      \end{equation*}
       \item If $ (p,q)\in (1,-1) $ and $ r\mid p-1 $ and ($ r\mid 3q-1 $ or $ r\mid 2q-4 $) and $ r\ne 5$ \textbf{OR} $ (p,q)\in (-1,1) $ and $ r\mid p+1 $ and ($ r\mid 3q+1 $ or $ r\mid 2q+4 $) and $ r\ne 5$, then
          \begin{equation*}
          LC(s)=\frac{3pq+q-2}{2}.
           \end{equation*}
       \item If $ (p,q)\in (1,-1) $ and $ r\mid p-1 $ and ($ r\mid 3q-1 $ or $ r\mid 2q-4 $) and $ r= 5$ \textbf{OR} $ (p,q)\in (-1,1) $ and $ r\mid p+1 $ and ($ r\mid 3q+1 $ or $ r\mid 2q+4 $) and $ r= 5$, then
            \begin{equation*}
              LC(s)=\frac{3pq+q-4}{2}.
            \end{equation*}
            \end{enumerate}
     \end{sec3thm1x}
     \begin{sec3_remark2}\label{Lab_sec3_remark2_A_thm2}
     In Theorem \ref{lab_MainThorem_01x}, if $ p\equiv 1\pmod 8 $ and $ r\mid p-1 $, by Lemma \ref{lab_sec3_lemma8} (1), $ r\mid 3q-1 $ and $ r\mid 2q-4 $ imply that $ \beta^{0}=1 $ and $ \beta^{pq} $ are zeros of  $ G_{s}(\beta^{k}) $ for $ k\in Z_{2T} $, respectively. For $ p\equiv -1\pmod 8 $ and $ r\mid p+1 $, again by Lemma \ref{lab_sec3_lemma8} (1), $ \beta^{0} $ is zero of $ G_{s}(\beta^{k}) $ if only if  $ r\mid 3q+1 $ and, $ \beta^{pq} $ is zero of $ G_{s}(\beta^{k}) $ if only if $ r\mid 2q+4 $. It is obvious that $ r=5 $ if only if both $ r\mid 3q-1 $ and $ r\mid 2q-4 $ hold for the case $ p\equiv 1\pmod 8 $ and $ r\mid p-1 $, or if only if both $ r\mid 3q+1 $ and $ r\mid 2q+4 $ hold for the case $ p\equiv -1\pmod 8 $ and $ r\mid p+1 $
     \end{sec3_remark2}
     \begin{sec3_remark3}\label{Lab_sec3_remark3_A_thm2}
     In Table \ref{Lab_Table3_lemma8_2case2}, les values of $ G_{s}(\beta^{k}) $ for $ k\in 2D_{0}^{(T)}\cup 2D_{1}^{(T)} $ that correspond to the eight sub-cases in Lemma \ref{lab_sec3_lemma8} (2) are listed, where $ A_{0}, A_{1}\in \lbrace 0,\;-1\rbrace $ and,  $ B_{0} $ and $ B_{1} $ take arbitrary values but $ B_{0} + B_{1}=-1$. If $ A_{0}=0 $, $ G_{s}(\beta^{k})=0 $ for the sub-cases $ d $ and $ h $ in Lemma \ref{lab_sec3_lemma8} (2); if $ A_{0}=-1 $, $ G_{s}(\beta^{k})=0 $ for the sub-cases $ b $ and $ h $ in Lemma \ref{lab_sec3_lemma8} (2). For above two cases $ A_{0}=0 $ and $ A_{0}=-1 $, from Lemma \ref{lab_sec3_lemma8} (2), we can observe that the condition $ \bigl(\tfrac{q}{p}\bigr) =\bigl(\tfrac{p}{q}\bigr) $ is required, that is equivalent to the condition $ p\equiv 1\pmod 8 $ and $ q\equiv -1\pmod 8 $ or $ p\equiv -1\pmod 8 $ and $ q\equiv 1\pmod 8 $ by the Quadratic Reciprocity law (see Lemma \ref{lab_sec2_lemma3x} (2)). In other words, for $ (p,q)\in (1,-1) $ or $ (p,q)\in (-1,1) $, if $ A_{0}, A_{1}\in \lbrace 0,\;-1\rbrace $, then $ G_{s}(\beta^{k}) $ for $ k\in 2D_{0}^{(T)}\cup 2D_{1}^{(T)} $, has $ \tfrac{(p-1)(q-1)}{2} $ zeros. But for $ (p,q)\in (-1,-1) $, $ G_{s}(\beta^{k}) $ for $ k\in 2D_{0}^{(T)}\cup 2D_{1}^{(T)} $ has no zeros.
     \end{sec3_remark3}
      \begin{table}[!t]
      \begin{threeparttable}[t]
        \renewcommand{\arraystretch}{1.3}
        \caption{Values of $ G_{s}(\beta^{k})$ in Lemma \ref{lab_sec3_lemma8}(2)\tnote{1}
        \label{Lab_Table3_lemma8_2case2}}
        \centering
        \begin{tabular}{c|c|c|c|c|c|c|c|c|c|c|c}
        \hline
       $ A_{0} $ & $ A_{1} $ & $ B_{0} $ & $ B_{1} $ & $ a $ & $ b $ & $ c $ & $ d $ & $ e $ & $ f $ & $ g $ & $ h $\\
        \hline
        0 & -1 & *\tnote{2} & * & * & * & * & 0 & * & * & * & 0\\
        \hline
        -1 & 0 & * & * & * &  0 & * & * & * & * & * & 0\\
        \hline
        \end{tabular}
         \begin{tablenotes}
              \item [1] $ \bigl(\tfrac{r}{p}\bigr) =1 $ and $ \bigl(\tfrac{r}{q}\bigr) =-1 $.
              \item [2] * means an arbitrary values.
             
              \end{tablenotes}
        \end{threeparttable}
        \end{table}
     \begin{proof}[Proof of Theorem \ref{lab_MainThorem_01x}]
     Since $ \bigl(\tfrac{r}{p}\bigr) =1 $ and $ \bigl(\tfrac{r}{q}\bigr) =-1 $, by Lemma \ref{lab_sec2_lemma4x}, $ A_{0}, A_{1}\in \mathbf{GF}(r)$ and $ B_{0}, B_{1}\in \mathbf{GF}(r^{m})\setminus \mathbf{GF}(r)$. If $ p\equiv 1 \pmod  8 $ and $ r\mid p-1 $ or $ p\equiv -1 \pmod  8 $ and $ r\mid p+1 $, then by Lemma \ref{lab_sec2_lemma1x}, $ A_{0}, A_{1}\in \lbrace 0,\;-1\rbrace $. From Lemma \ref{lab_sec3_lemma8} (3), $ G_{s}(\beta^{k}) $ for $ k\in 2qD_{0}^{(p)}\cup 2qD_{1}^{(p)} $ has $ \tfrac{p-1}{2} $ zeros. We next prove each case in Theorem \ref{lab_MainThorem_01x}.
     \begin{enumerate}
     \item Case $ (p,q)\in (1,-1) $ and $ r\mid p-1 $ and $ r\nmid 3q-1 $ and $ r\nmid 2q-4 $ \textbf{OR} $ (p,q)\in (-1,1) $ and $ r\mid p+1 $ and $ r\nmid 3q+1 $ and $ r\nmid 2q+4 $. From Remark \ref{Lab_sec3_remark2_A_thm2}, neither $ \beta^{0}=1 $ nor $ \beta^{pq} $ is a zero of $ G_{s}(\beta^{k}) $ for $ k\in \lbrace 0,\ pq \rbrace \cup Z^{*}_{2T}\cup qZ^{*}_{2p}\cup pZ^{*}_{2q} $. From Remark \ref{Lab_sec3_remark3_A_thm2}, $ G_{s}(\beta^{k}) $ for $ k\in 2D_{0}^{(T)}\cup 2D_{1}^{(T)} $ has $ \tfrac{(p-1)(q-1)}{2} $ zeros. In addition that there are $ \tfrac{p-1}{2} $ zeros for $ k\in 2qD_{0}^{(p)}\cup 2qD_{1}^{(p)} $, the linear complexity could be directly computed out by 
     \begin{equation*}
     LC(s)=2pq-\frac{(p-1)(q-1)}{2}-\frac{p-1}{2}=\frac{3pq+q}{2}.
     \end{equation*}
     \item Case $ (p,q)\in (1,-1) $ and $ r\mid p-1 $ and ($ r\mid 3q-1 $ or $ r\mid 2q-4 $) and $ r\ne 5$ \textbf{OR} $ (p,q)\in (-1,1) $ and $ r\mid p+1 $ and ($ r\mid 3q+1 $ or $ r\mid 2q+4 $) and $ r\ne 5$. By above discussion, these conditions imply that $ G_{s}(\beta^{k}) $ has an extra zero that is either $ \beta^{0}=1 $ or $ \beta^{pq} $, so the linear complexity equals
        \begin{equation*}
        LC(s)=2pq-\frac{(p-1)(q-1)}{2}-\frac{p-1}{2}-1=\frac{3pq+q-2}{2}.
        \end{equation*}
     \item Case $ (p,q)\in (1,-1) $ and $ r\mid p-1 $ and ($ r\mid 3q-1 $ or $ r\mid 2q-4 $) and $ r= 5$ \textbf{OR} $ (p,q)\in (-1,1) $ and $ r\mid p+1 $ and ($ r\mid 3q+1 $ or $ r\mid 2q+4 $) and $ r= 5$. These conditions imply that $ G_{s}(\beta^{k}) $ has two extra zeros that are  $ \beta^{0}=1 $ and $ \beta^{pq} $, so the linear complexity equals
           \begin{equation*}
           LC(s)=2pq-\frac{(p-1)(q-1)}{2}-\frac{p-1}{2}-2=\frac{3pq+q-4}{2}.
           \end{equation*}
     \end{enumerate}
     \end{proof}
     
       \begin{sec3thm1xx}\label{lab_MainThorem_01xx}
       Let $ (2,2) \in  D_{0}^{(p)}\times D_{0}^{(q)}$, $ \bigl(\tfrac{r}{p}\bigr) =1 $, $ \bigl(\tfrac{r}{q}\bigr) =-1 $, and $ (p,q)\in (-1,-1) $. Then, the linear complexity over $ \mathbf{GF}(r^{m}) $ of the generalized cyclotomic quaternary sequences with period $ 2T $ given in (\ref{lab_seq_chang_ex}), $  LC(s) $ can be determined according to following cases.
           \begin{enumerate} 
           \item If   $ r\mid p+1 $ and $ r\nmid 3q+1 $ and $ r\nmid 2q+4 $, then
           \begin{equation*}
              LC(s)=\frac{4pq+p-1}{2}.
           \end{equation*}
            \item If  $ r\mid p+1 $ and ($ r\mid 3q+1 $ or $ r\mid 2q+4 $) and $ r\ne 5$, then
               \begin{equation*}
               LC(s)=\frac{4pq+p-3}{2}.
                \end{equation*}
            \item If  $ r\mid p+1 $ and ($ r\mid 3q+1 $ or $ r\mid 2q+4 $) and $ r= 5$, then
                 \begin{equation*}
                   LC(s)=\frac{4pq+p-5}{2}.
                 \end{equation*}
                 \end{enumerate}
       \end{sec3thm1xx} 
        \begin{proof}
        Since $ (p,q)\in (-1,-1) $, by Remark \ref{Lab_sec3_remark3_A_thm2}, $ G_{s}(\beta^{k}) $ for $ k\in 2D_{0}^{(T)}\cup 2D_{1}^{(T)} $ has no zeros. The rest of the proof is similar to that for Theorem \ref{lab_MainThorem_01x}.
        \end{proof}
 \subsection{$ 2 $ is a quadratic non-residue both modulo $ p $ and modulo $ q $}  
    
  \begin{sec3lemma9}\label{lab_sec3_lemma9}
  Let $ (2,\ 2)\in D_{1}^{(p)}\times D_{1}^{(q)}$. Then the values of $ G_{s}(\beta^{k}) $ for $ k\in Z_{2T} $ can be determined as follows:
  \begin{enumerate} 
  \item $ k\in \lbrace 0,\ pq \rbrace \cup 2qZ^{*}_{p}\cup 2pZ^{*}_{q} $.
  \begin{equation*}
  G_{s}(\beta^{k})=
  \begin{cases}
  3pq-1&k\in\lbrace 0\rbrace,\\
  2pq-4&k\in\lbrace pq\rbrace,\\
  -1   &k\in 2qZ^{*}_{p}\cup 2pZ^{*}_{q}.
  \end{cases}
  \end{equation*}
  \item $ k\in D_{0}^{(2T)}\cup D_{1}^{(2T)} $. Let  $ k_{1}\equiv k\pmod p $ and $ k_{2}\equiv k\pmod q $.
  \begin{enumerate}
  \item
  If $ k\in D_{0}^{(2T)} $ and $ (k_{1},\ k_{2})\in D^{(p)}_{0}\times D^{(q)}_{0} $, then
  \begin{equation*}
  G_{s}(\beta^{k})=A_{0}-A_{1}+B_{0}-B_{1}-4.
  \end{equation*} 
  \item
  If $ k\in D_{0}^{(2T)} $ and $ (k_{1},\ k_{2})\in D^{(p)}_{1}\times D^{(q)}_{1} $, then
  \begin{equation*}
  G_{s}(\beta^{k})=A_{1}-A_{0}+B_{1}-B_{0}-4.
  \end{equation*}
  \item
  If $ k\in D_{1}^{(2T)} $ and $ (k_{1},\ k_{2})\in D^{(p)}_{0}\times D^{(q)}_{1} $, then
  \begin{equation*}
  G_{s}(\beta^{k})=A_{0}-A_{1}+B_{1}-B_{0}-4.
  \end{equation*} 
  \item
  If $ k\in D_{1}^{(2T)} $ and $ (k_{1},\ k_{2})\in D^{(p)}_{1}\times D^{(q)}_{0} $, then
  \begin{equation*}
  G_{s}(\beta^{k})=A_{1}-A_{0}+B_{0}-B_{1}-4.
  \end{equation*}  
  \end{enumerate}
  \item $ k\in 2D_{0}^{(T)}\cup 2D_{1}^{(T)} $.
  \begin{enumerate}
  \item
  If $ k\in 2D_{0}^{(T)} $ and $ \biggl(\frac{q}{p}\biggr)=-\biggl(\frac{p}{q}\biggr) $ or $ k\in 2D_{1}^{(T)} $ and $ \biggl(\frac{q}{p}\biggr)=\biggl(\frac{p}{q}\biggr) $, then
  \begin{equation*}
  G_{s}(\beta^{k})=2(A_{0}B_{0}+A_{1}B_{1}-1).
  \end{equation*}
  \item
  If $ k\in 2D_{0}^{(T)} $ and $ \biggl(\frac{q}{p}\biggr)=\biggl(\frac{p}{q}\biggr) $ or $ k\in 2D_{1}^{(T)} $ and $ \biggl(\frac{q}{p}\biggr)=-\biggl(\frac{p}{q}\biggr) $, then
  \begin{equation*}
  G_{s}(\beta^{k})=2(A_{0}B_{1}+A_{1}B_{0}-1).
  \end{equation*}
  \end{enumerate}
  \item $ k\in qD_{0}^{(2p)}\cup qD_{1}^{(2p)} $.
  \begin{enumerate}
  \item
  If $ k\in qD_{0}^{(2p)} $ and $ \biggl(\frac{q}{p}\biggr)=1 $ or $ k\in qD_{1}^{(2p)} $ and $ \biggl(\frac{q}{p}\biggr)=-1 $, then
  \begin{equation*}
  G_{s}(\beta^{k})=A_{1}-A_{0}-4.
  \end{equation*}
  \item
  If $ k\in qD_{0}^{(2p)} $ and $ \biggl(\frac{q}{p}\biggr)=-1 $ or $ k\in qD_{1}^{(2p)} $ and $ \biggl(\frac{q}{p}\biggr)=1 $, then
  \begin{equation*}
  G_{s}(\beta^{k})=A_{0}-A_{1}-4.
  \end{equation*}
  \end{enumerate}
  \item $ k\in pD_{0}^{(2q)}\cup pD_{1}^{(2q)} $.
  \begin{enumerate}
  \item
  If $ k\in pD_{0}^{(2q)} $ and $ \biggl(\frac{p}{q}\biggr)=1 $ or $ k\in pD_{1}^{(2q)} $ and $ \biggl(\frac{p}{q}\biggr)=-1 $, then
  \begin{equation*}
  G_{s}(\beta^{k})=B_{1}-B_{0}-4.
  \end{equation*}
  \item
  If $ k\in pD_{0}^{(2q)} $ and $ \biggl(\frac{p}{q}\biggr)=-1 $ or $ k\in pD_{1}^{(2q)} $ and $ \biggl(\frac{p}{q}\biggr)=1 $, then
  \begin{equation*}
  G_{s}(\beta^{k})=B_{0}-B_{1}-4.
  \end{equation*}
  \end{enumerate}
  \end{enumerate}
  \end{sec3lemma9}
  \begin{proof}
  It is clear that $ 2_{p}^{-1}\in D_{1}^{(p)} $ and $ 2_{q}^{-1}\in D_{1}^{(q)} $. By Lemma \ref{lab_sec3_lemma1} and Lemma \ref{lab_sec3_lemma4}, $ 2_{T}^{-1}\in D_{0}^{(T)} $. By Lemma \ref{lab_sec2_lemma2x}, it can be deduced the following equations
  \begin{equation*}
  \begin{split}
  &\sum_{i\in D_{1}^{(T)}}\beta_{T}^{2_{T}^{-1}ki}=\sum_{i\in 2_{T}^{-1}D_{1}^{(T)}}\beta_{T}^{ki}=\sum_{i\in D_{1}^{(T)}}\beta_{T}^{ki},\\
  &\sum_{i\in D_{1}^{(p)}}\beta_{p}^{2_{p}^{-1}ki}=\sum_{i\in 2_{p}^{-1}D_{1}^{(p)}}\beta_{p}^{ki}=\sum_{i\in D_{0}^{(p)}}\beta_{p}^{ki},\\
  &\sum_{i\in D_{1}^{(q)}}\beta_{q}^{2_{q}^{-1}ki}=\sum_{i\in 2_{q}^{-1}D_{1}^{(q)}}\beta_{q}^{ki}=\sum_{i\in D_{0}^{(q)}}\beta_{q}^{ki}.
  \end{split}
  \end{equation*}
  Substitute the group of above equations into (\ref{lab_seq_chang_GP_in_beta_k}), it leads to a new form for $ G_{s}(\beta^{k}) $ with $ k\in Z_{2T} $:
  \begin{equation}\label{Lab_Lemma9_prof_Gsbetak}
  \begin{split}
  G_{s}(\beta^{k})=2\beta_{2}^{k}&+\biggl(\beta_{2}^{k}+1\biggr)\sum_{i\in D_{1}^{(T)}}\beta_{T}^{ki}\\
     &+\biggl(\beta_{2}^{k}\sum_{i\in D_{0}^{(p)}}\beta_{p}^{ki}+\sum_{i\in D_{1}^{(p)}}\beta_{p}^{ki}\biggr)\\
     &+\biggl(\beta_{2}^{k}\sum_{i\in D_{0}^{(q)}}\beta_{q}^{ki}+\sum_{i\in D_{1}^{(q)}}\beta_{q}^{ki}\biggr)\\
     &+2\biggl(Z^{(p)}(k)\cdot Z^{(q)}(k)+Z^{(p)}(k)+Z^{(q)}(k)\biggr).
  \end{split}
  \end{equation}
  The rest of the proof is similar to that of Lemma \ref{lab_sec3_lemma8}.
  \end{proof}  
  
   \begin{sec3Bcor1}\label{lab_sec3Bcor1}
   Let $ 2\in D_{1}^{(p)} $, $ 2\in D_{1}^{(q)} $, $ \bigl(\tfrac{r}{p}\bigr) =1$ and $ \bigl(\tfrac{r}{q}\bigr) =1$. In addition, let $ r\mid p+16 $ and $ r\mid q+16 $ if $ p\equiv 3 \pmod 8 $ and $ q\equiv 3 \pmod 8 $, $ r\mid p-16 $ and $ r\mid q+16 $ if $ p\equiv 5 \pmod 8 $ and $ q\equiv 3 \pmod 8 $ and, $ r\mid p+16 $ and $ r\mid q-16 $ if $ p\equiv 3 \pmod 8 $ and $ q\equiv 5 \pmod 8 $. Then, $ G_{s}(\beta^{k}) $, where  $ k\in qD_{0}^{(2p)}\cup  qD_{1}^{(2p)}\cup pD_{0}^{(2q)}\cup  pD_{1}^{(2q)} $, has exactly $ \tfrac{p-1}{2}+\tfrac{q-1}{2} $'s zeros.
   \end{sec3Bcor1}
   \begin{proof}
   Since $ \bigl(\tfrac{r}{p}\bigr) =1$ and $ \bigl(\tfrac{r}{q}\bigr) =1$, by Lemma \ref{lab_sec2_lemma4x}, $ A_{0},A_{1},B_{0},B_{1} \in \mathbf{GF}(r)$. Using Lemma \ref{lab_sec2_lemma1x}, for the case where $ r\mid p+16 $ and $ r\mid q+16 $ with $ p\equiv 3 \pmod 8 $ and $ q\equiv 3 \pmod 8 $, resolve next congruent system:
   \begin{equation}\label{lab_sec3Bcor1_eq1}
   \begin{cases}
   A_{0}+A_{1}&=-1,\\
   B_{0}+B_{1}&=-1,\\
   p+16&\equiv 0 \pmod r,\\
   q+16&\equiv 0 \pmod r,\\
   A_{1}(A_{1}+1)&\equiv -\frac{p+1}{4} \pmod r,\\
   B_{1}(B_{1}+1)&\equiv -\frac{q+1}{4} \pmod r.
   \end{cases}
   \end{equation}
    Now consider the case where $ r\mid p-16 $ and $ r\mid q+16 $ with $ p\equiv -3 \pmod 8 $ and $ q\equiv 3 \pmod 8 $ and next congruent system:  
    \begin{equation}\label{lab_sec3Bcor1_eq2}
     \begin{cases}
     A_{0}+A_{1}&=-1,\\
     B_{0}+B_{1}&=-1,\\
     p-16&\equiv 0 \pmod r,\\
     q+16&\equiv 0 \pmod r,\\
     A_{0}(A_{0}+1)&\equiv \frac{p-1}{4} \pmod r,\\
     B_{1}(B_{1}+1)&\equiv -\frac{q+1}{4} \pmod r.
     \end{cases}
     \end{equation}
   And finally for the case where $ r\mid p+16 $ and $ r\mid q-16 $ with $ p\equiv 3 \pmod 8 $ and $ q\equiv -3 \pmod 8 $, consider the following  congruent system: 
    \begin{equation}\label{lab_sec3Bcor1_eq3}
    \begin{cases}
    A_{0}+A_{1}&=-1,\\
    B_{0}+B_{1}&=-1,\\
    p+16&\equiv 0 \pmod r,\\
    q-16&\equiv 0 \pmod r,\\
    A_{1}(A_{1}+1)&\equiv -\frac{p+1}{4} \pmod r,\\
    B_{0}(B_{0}+1)&\equiv \frac{q-1}{4} \pmod r.
    \end{cases}
    \end{equation}
    By a simple modular arithmetic computation, the congruent systems (\ref{lab_sec3Bcor1_eq1}), (\ref{lab_sec3Bcor1_eq2}) and (\ref{lab_sec3Bcor1_eq3}) give rise to the same solutions:
    \begin{equation}\label{lab_sec3Bcor1_eq4_solA}
     \pm(A_{0}-A_{1})-4\equiv 0 \pmod r,
    \end{equation}
    and
    \begin{equation}\label{lab_sec3Bcor1_eq4_solB}
       \pm(B_{0}-B_{1})-4\equiv 0 \pmod r.
    \end{equation}
    If $ A_{0}-A_{1}-4\equiv 0 \pmod r $ or $ A_{1}-A_{0}-4\equiv 0 \pmod r $, then by Lemma \ref{lab_sec3_lemma9} (4), $ G_{s}(\beta^{k}) $, where $ k\in qD_{0}^{(2p)}\cup  qD_{1}^{(2p)}$, has $ \tfrac{p-1}{2} $'zeros. On the other hand, if $ B_{0}-B_{1}-4\equiv 0 \pmod r $ or $ B_{1}-B_{0}-4\equiv 0 \pmod r $, then by Lemma \ref{lab_sec3_lemma9} (5), $ G_{s}(\beta^{k}) $, where $ k\in pD_{0}^{(2q)}\cup  pD_{1}^{(2q)}$, has $ \tfrac{q-1}{2} $'zeros. From above discussion, (\ref{lab_sec3Bcor1_eq4_solA}) and (\ref{lab_sec3Bcor1_eq4_solB}), $ G_{s}(\beta^{k}) $ has exactly $ \tfrac{p-1}{2}+\tfrac{q-1}{2} $'s zeros, where $ k\in qD_{0}^{(2p)}\cup  qD_{1}^{(2p)}\cup pD_{0}^{(2q)}\cup  pD_{1}^{(2q)} $.
   \end{proof}
    \begin{sec3Bcor2}\label{lab_sec3Bcor2}
    Let $ 2\in D_{1}^{(p)} $, $ 2\in D_{1}^{(q)} $, $ \bigl(\tfrac{r}{p}\bigr) =1$ and $ \bigl(\tfrac{r}{q}\bigr) =1$. In addition, let $ r\mid p+1 $ and $ r\mid q+1 $ if $ p\equiv 3 \pmod 8 $ and $ q\equiv 3 \pmod 8 $, $ r\mid p-1 $ and $ r\mid q+1 $ if $ p\equiv 5 \pmod 8 $ and $ q\equiv 3 \pmod 8 $ and, $ r\mid p+1 $ and $ r\mid q-1 $ if $ p\equiv 3 \pmod 8 $ and $ q\equiv 5 \pmod 8 $. Then, $ G_{s}(\beta^{k}) $ has exactly $ \tfrac{(p-1)(q-1)}{2} $'s zeros, where $ k\in 2D_{0}^{(T)}\cup 2D_{1}^{(T)} $.
    \end{sec3Bcor2}
   \begin{proof}
   By Lemma \ref{lab_sec2_lemma4x}, $ A_{0},A_{1},B_{0},B_{1} \in \mathbf{GF}(r)$. By Lemma \ref{lab_sec2_lemma1x}, $ A_{0},A_{1},B_{0},B_{1} \in \lbrace 0,-1\rbrace$, since $ p,q\equiv\pm 1\pmod 4 $. By Lemma \ref{lab_sec3_lemma9} (3), $ A_{0}B_{0}+A_{1}B_{1}=1 $ or $ A_{0}B_{1}+A_{1}B_{0}=1 $. In other words,  $ G_{s}(\beta^{k}) $ has  $ \tfrac{(p-1)(q-1)}{2} $'s zeros, where $ k\in 2D_{0}^{(T)}\cup 2D_{1}^{(T)} $.
   \end{proof}
    \begin{sec3Bcor3}\label{lab_sec3Bcor3}
    Let $ 2\in D_{1}^{(p)} $, $ 2\in D_{1}^{(q)} $, $ \bigl(\tfrac{r}{p}\bigr) =1$ and $ \bigl(\tfrac{r}{q}\bigr) =1$. In addition, let $ r\mid p+16 $ and $ r\mid q+16 $ if $ p\equiv 3 \pmod 8 $ and $ q\equiv 3 \pmod 8 $, $ r\mid p-16 $ and $ r\mid q+16 $ if $ p\equiv 5 \pmod 8 $ and $ q\equiv 3 \pmod 8 $ and, $ r\mid p+16 $ and $ r\mid q-16 $ if $ p\equiv 3 \pmod 8 $ and $ q\equiv 5 \pmod 8 $. If $ r=5 $, then $ G_{s}(\beta^{k}) $ has at least $ \tfrac{(p-1)(q-1)}{2}+\tfrac{p-1}{2}+\tfrac{q-1}{2} $'s zeros, where $ k\in Z_{2T} $.
    \end{sec3Bcor3}
    \begin{proof}
    Remark that $ 5\mid p\pm 16 $ and $ 5\mid q\pm 16 $ imply $ 5\mid p\pm 1 $ and $ 5\mid q\pm 1 $. By Corollary \ref{lab_sec3Bcor1} and \ref{lab_sec3Bcor2}, it follows the result of Corollary \ref{lab_sec3Bcor3}.
    \end{proof}
    \begin{sec3Bcor4}\label{lab_sec3Bcor4}
      Let $ 2\in D_{1}^{(p)} $, $ 2\in D_{1}^{(q)} $, $ \bigl(\tfrac{r}{p}\bigr) =1$ and $ \bigl(\tfrac{r}{q}\bigr) =1$. In addition, let $ r\mid p+4 $ and $ r\mid q+4 $ if $ p\equiv 3 \pmod 8 $ and $ q\equiv 3 \pmod 8 $, $ r\mid p-4 $ and $ r\mid q+4 $ if $ p\equiv 5 \pmod 8 $ and $ q\equiv 3 \pmod 8 $ and, $ r\mid p+4 $ and $ r\mid q-4 $ if $ p\equiv 3 \pmod 8 $ and $ q\equiv 5 \pmod 8 $.  Then $ G_{s}(\beta^{k}) $ has  $ \tfrac{(p-1)(q-1)}{4} $'s zeros, where $ k\in D_{0}^{(2T)}\cup D_{1}^{(2T)} $
      \end{sec3Bcor4}
      \begin{proof}
      By Lemma \ref{lab_sec2_lemma4x}, $ A_{0},A_{1},B_{0},B_{1} \in \mathbf{GF}(r)$. By Lemma \ref{lab_sec2_lemma1x} and using the same proof method as for Corollary \ref{lab_sec3Bcor1}, it can be deduced the following equations:
      \begin{equation*}
      \begin{cases}
      \pm( A_{0}- A_{1}) &\equiv 2 \pmod r,\\
      \pm( B_{0}- B_{1}) &\equiv 2 \pmod r.
      \end{cases}
      \end{equation*} 
      Substitute the above equations into Lemma \ref{lab_sec3_lemma9} (2), it leads to that in one of four sub-cases, $ G_{s}(\beta^{k})=0 $. In other words, there are $ \tfrac{(p-1)(q-1)}{4} $'s zeros for $ G_{s}(\beta^{k})$ where $ k\in D_{0}^{(2T)}\cup D_{1}^{(2T)} $.
      \end{proof} 
       \begin{sec3thm2}\label{lab_MainThorem_02}
         Let $ 2\in D_{1}^{(p)} $, $ 2\in D_{1}^{(q)} $, $ \bigl(\tfrac{r}{p}\bigr) =1$ and $ \bigl(\tfrac{r}{q}\bigr) =1$. In addition, let $ r\mid p+16 $ and $ r\mid q+16 $ if $ p\equiv 3 \pmod 8 $ and $ q\equiv 3 \pmod 8 $, $ r\mid p-16 $ and $ r\mid q+16 $ if $ p\equiv 5 \pmod 8 $ and $ q\equiv 3 \pmod 8 $ and, $ r\mid p+16 $ and $ r\mid q-16 $ if $ p\equiv 3 \pmod 8 $ and $ q\equiv 5 \pmod 8 $. Then, the linear complexity over $ \mathbf{GF}(r^{m}) $ of the generalized cyclotomic quaternary sequences with period $ 2T $ specified in (\ref{lab_seq_chang_ex}) can be determined according to following cases:
          \begin{enumerate} 
          \item if $ (p,q)\in (3,3) $ and $ r\notin\left\lbrace 5,13,59,127\right\rbrace $ \textbf{OR} $ (p,q)\in (3,-3) \cup (-3,3)$ and $ r\notin\left\lbrace 5,43,769\right\rbrace $.
          \begin{equation*}
          LC(s)=2pq-\dfrac{p+q}{2}+1
          \end{equation*}
          \item If $ (p,q)\in (3,3) $ and $ r\in\left\lbrace 13,59,127\right\rbrace $ \textbf{OR} $ (p,q)\in (3,-3) \cup (-3,3)$ and  $ r\in\left\lbrace 43,769\right\rbrace $.
          \begin{equation*}
           LC(s)=2pq-\dfrac{p+q}{2}
           \end{equation*}
           \item If $ r=5 $, then 
           \begin{equation*}
             LC(s)=\dfrac{3pq+1}{2}
             \end{equation*}
      \end{enumerate}
      \end{sec3thm2}
      
      \begin{sec3_B_remark1}\label{Lab_sec3_B_remark1}
      Follow the context in Theorem \ref{lab_MainThorem_02}, and let  $ r\mid p+16 $ and $ r\mid q+16 $ for the case where $ p\equiv 3 \pmod 8 $ and $ q\equiv 3 \pmod 8 $. It is easy to check that if $ r\in\left\lbrace 13,59\right\rbrace $ then $ r\mid 3pq-1 $, and if $ r=127 $ then $ r\mid 2pq-4 $. By Lemma \ref{lab_sec3_lemma9} (1), $ r\mid 3pq-1 $ implies  that $ \beta^{0}=1 $ is a zero of $ G_{s}(\beta^{k}) $, and $ r\mid 2pq-4 $ leads to that $ \beta^{pq}$ is a zero of $ G_{s}(\beta^{k}) $ as well, where $ k\in Z_{2T} $.
      \end{sec3_B_remark1}
      
      \begin{sec3_B_remark2}\label{Lab_sec3_B_remark2}
      Consider the context in Theorem \ref{lab_MainThorem_02}. Let $ r\mid p-16 $ and $ r\mid q+16 $ if $ p\equiv 5 \pmod 8 $ and $ q\equiv 3 \pmod 8 $, and $ r\mid p+16 $ and $ r\mid q-16 $ if $ p\equiv 3 \pmod 8 $ and $ q\equiv 5 \pmod 8 $. It is straightforward to verify that if $ r=769 $ then $ r\mid 3pq-1 $, and if $ r=43 $ then $ r\mid 2pq-4 $. Combining Remark \ref{Lab_sec3_B_remark1},  \ref{Lab_sec3_B_remark2}, and the condition  $ r\notin\left\lbrace 5,13,59,127\right\rbrace $ or $ r\notin\left\lbrace 5,43,769\right\rbrace $, it prevents that $ \beta^{0}=1 $ and $ \beta^{pq}$ are zeros to $ G_{s}(\beta^{k}) $, where $ k\in Z_{2T} $, and the case where $ r=5 $.
      \end{sec3_B_remark2}
      
      \begin{proof}[Proof of Theorem \ref{lab_MainThorem_02}]
      We prove each case of Theorem \ref{lab_MainThorem_02}.
      \begin{enumerate}
      \item By Remark \ref{Lab_sec3_B_remark2} and Corollary \ref{lab_sec3Bcor1}, $ G_{s}(\beta^{k}) $ has exactly $ \tfrac{p-1}{2}+\tfrac{q-1}{2} $'s zeros, where $ k\in Z_{2T} $. By (\ref{lab_eq_GCD_LC_eq_ex}), 
      \begin{equation*}
      LC(s)=2pq-\frac{p-1}{2}-\frac{q-1}{2}=2pq-\frac{p+q}{2}+1.
      \end{equation*}
      \item By Remark \ref{Lab_sec3_B_remark1}, either $ \beta^{0}=1 $ or $ \beta^{pq}$ but not both is a zero. Hence, again by Corollary \ref{lab_sec3Bcor1}, $ G_{s}(\beta^{k}) $ has exactly $ \tfrac{p-1}{2}+\tfrac{q-1}{2}+1 $'s zeros, where $ k\in Z_{2T} $. Hence,
      \begin{equation*}
      LC(s)=2pq-\frac{p-1}{2}-\frac{q-1}{2}-1=2pq-\frac{p+q}{2}.
      \end{equation*}
      \item By Corollary \ref{lab_sec3Bcor3}, $G_{s}(\beta^{k}) $ has exactly $ \tfrac{p-1}{2}+\tfrac{q-1}{2}+\tfrac{(p-1)(q-1)}{2}$'s zeros, where $ k\in Z_{2T} $. By (\ref{lab_eq_GCD_LC_eq_ex}), 
      \begin{equation*}
      \begin{split}
      LC(s)&=2pq-\frac{p-1}{2}-\frac{q-1}{2}-\frac{(p-1)(q-1)}{2}\\
           &=\frac{3pq+1}{2}.
      \end{split}
      \end{equation*}
      \end{enumerate}
      \end{proof}
    
 \begin{sec3thm2aux1}\label{lab_sec3thm2aux1}
  Let $ 2\in D_{1}^{(p)} $, $ 2\in D_{1}^{(q)} $, $ \bigl(\tfrac{r}{p}\bigr) =1$ and $ \bigl(\tfrac{r}{q}\bigr) =1$. In addition, let $ r\mid p+4 $ and $ r\mid q+4 $ if $ p\equiv 3 \pmod 8 $ and $ q\equiv 3 \pmod 8 $, $ r\mid p-4 $ and $ r\mid q+4 $ if $ p\equiv 5 \pmod 8 $ and $ q\equiv 3 \pmod 8 $ and, $ r\mid p+4 $ and $ r\mid q-4 $ if $ p\equiv 3 \pmod 8 $ and $ q\equiv 5 \pmod 8 $. Then, the linear complexity over $ \mathbf{GF}(r^{m}) $ of the generalized cyclotomic quaternary sequences with period $ 2T $ specified in (\ref{lab_seq_chang_ex}) can be determined according to following cases:
  \begin{enumerate} 
        \item if $ (p,q)\in (3,3) $ and  $ r\notin\left\lbrace 7,47\right\rbrace $ \textbf{OR} $ (p,q)\in (3,-3) \cup (-3,3)$ and $ r\ne 7 $, then
    \begin{equation*}
        LC(s)=\dfrac{7pq+p+q-1}{4}
     \end{equation*}
     \item If $ (p,q)\in (3,3) $ and  $ r\in\left\lbrace 7,47\right\rbrace $ \textbf{OR} $ (p,q)\in (3,-3) \cup (-3,3)$ and $ r=7 $, then
     \begin{equation*}
      LC(s)=\dfrac{7pq+p+q-5}{4}
      \end{equation*}
  \end{enumerate}
  \end{sec3thm2aux1}
  
  \begin{sec3_B_remark3}\label{Lab_sec3_B_remark3}
  In Theorem \ref{lab_sec3thm2aux1}, for the case where $ (p,q)\in (3,3) $, if $ r=47 $, then $ r\mid 3pq-1 $, and if $ r=7 $, then $ r\mid 2pq-4 $. In  other words, if $ r\in \lbrace 7,47\rbrace $, then either $ \beta^{0} =1$ or $ \beta^{pq}  $ but not both is a zero of $ G_{s}(\beta^{k}) $ , using Lemma \ref{lab_sec3_lemma9} (1). For the case $ (p,q)\in (3,-3)\cup (-3,3) $, if $ r=7 $, then $ r\mid 3pq-1 $. That is, $ \beta^{0} =1$ is a zero of $ G_{s}(\beta^{k}) $. Where $ k\in Z_{2T} $
  \end{sec3_B_remark3}
  \begin{proof}[Proof of Theorem \ref{lab_sec3thm2aux1}]
  Using Remark \ref{Lab_sec3_B_remark3} and Corollary \ref{lab_sec3Bcor4}.
  \end{proof}
  
  \begin{sec3thm2aux2}\label{lab_sec3thm2aux2}
  Let $ 2\in D_{1}^{(p)} $, $ 2\in D_{1}^{(q)} $, $ \bigl(\tfrac{r}{p}\bigr) =1$ and $ \bigl(\tfrac{r}{q}\bigr) =1$. In addition, let $ r\mid p+1 $ and $ r\mid q+1 $ if $ p\equiv 3 \pmod 8 $ and $ q\equiv 3 \pmod 8 $, $ r\mid p-1 $ and $ r\mid q+1 $ if $ p\equiv 5 \pmod 8 $ and $ q\equiv 3 \pmod 8 $ and, $ r\mid p+1 $ and $ r\mid q-1 $ if $ p\equiv 3 \pmod 8 $ and $ q\equiv 5 \pmod 8 $. Then, the linear complexity over $ \mathbf{GF}(r^{m}) $ of the generalized cyclotomic quaternary sequences with period $ 2T $ specified in (\ref{lab_seq_chang_ex}) can be determined according to following cases:
  \begin{enumerate} 
  \item if $ (p,q)\in (3,3) $ and $ r\ne 5 $ \textbf{OR} $ (p,q)\in (3,-3) \cup (-3,3)$ and $ r\ne 5 $, then
      \begin{equation*}
       LC(s)=\dfrac{3pq+p+q-1}{2}
       \end{equation*}
       \item If $ r= 5 $, then 
       \begin{equation*}
          LC(s)=\dfrac{3pq+1}{2}
          \end{equation*}
    \end{enumerate}
    \end{sec3thm2aux2}
  \begin{proof}
   Using  Corollary \ref{lab_sec3Bcor2} and \ref{lab_sec3Bcor3}.
   \end{proof}  
     \begin{sec3thm2x}\label{lab_MainThorem_02x}
       Let $ 2\in D_{1}^{(p)} $, $ 2\in D_{1}^{(q)} $, $ \bigl(\tfrac{r}{p}\bigr) =1$ and $ \bigl(\tfrac{r}{q}\bigr) =-1$. In addition, let $ r\mid p+16 $ if $ p\equiv 3 \pmod 8 $ , $ r\mid p-16 $  if $ p\equiv 5 \pmod 8 $. Then, the linear complexity over $ \mathbf{GF}(r^{m}) $ of the generalized cyclotomic quaternary sequences with period $ 2T $ specified in (\ref{lab_seq_chang_ex}) can be determined according to following cases:
     \begin{enumerate} 
    \item if $ (p,q)\in (3,\pm 3) $ and $ r\nmid 48q+1 $ and $ r\nmid 8q+1 $ \textbf{OR } $ (p,q)\in (-3,3) $  and $ r\nmid 48q-1 $ and $ r\nmid 8q-1 $, then 
     \begin{equation*}
     LC(s)=2pq-\dfrac{p-1}{2}
     \end{equation*}
       \item If $ (p,q)\in (3,\pm 3) $ and ($ r\mid 48q+1 $ or $ r\mid 8q+1 $) and $ r\ne 5 $ \textbf{OR} $ (p,q)\in (-3,3) $ and ($ r\mid 48q-1 $ or $ r\mid 8q-1 $) and $ r\ne 5 $, then 
      \begin{equation*}
       LC(s)=2pq-\dfrac{p-1}{2}-1.
       \end{equation*}
         \item If $ (p,q)\in (3,\pm 3) $ and ($ r\mid 48q+1 $ or $ r\mid 8q+1 $) and $ r=5 $ \textbf{OR} $ (p,q)\in (-3,3) $ and ($ r\mid 48q-1 $ or $ r\mid 8q-1 $) and $ r= 5 $, then 
         \begin{equation*}
         LC(s)=2pq-\dfrac{p-1}{2}-2.
           \end{equation*}
       \end{enumerate}
       \end{sec3thm2x}
   \begin{sec3_B_remark4}\label{Lab_sec3_B_remark4}
   In Theorem \ref{lab_MainThorem_02x}, the condition $ r\mid 48q+1 $ or $ r\mid 8q+1 $ for the case where $ p\equiv 3\pmod 8 $ and $ r\mid p+16 $, and the one  $ r\mid 48q-1 $ or $ r\mid 8q-1 $ for the case where $ p\equiv 5\pmod 8 $ and $ r\mid p-16 $ imply that by Lemma \ref{lab_sec3_lemma9} (1), $ \beta^{0}=1 $ or $ \beta^{pq}$ is a zero of $ G_{s}(\beta^{k}) $ for $ k\in Z_{2T} $, respectively. It is obvious that if $ r=5 $, then $ r\mid 48q+1 $ implies $ r\mid 8q+1 $, vice versa, and $ r\mid 48q-1 $ implies $ r\mid 8q-1 $, vice versa.
   \end{sec3_B_remark4}
   \begin{proof}[Proof of Theorem \ref{lab_MainThorem_02x}]
     Using  Corollary \ref{lab_sec3Bcor1} and Remark \ref{Lab_sec3_B_remark4}.
   \end{proof}  
   \begin{sec3Bexp1}\label{lab_sec3Bexp1}
   We give some pairs of $ (p,q) $ which match the condition that is $ (p,q)\in (3, 3) $ and ($ r\mid 48q+1 $ or $ r\mid 8q+1 $) and $ r=5 $, and compute the linear complexity both by Algorithm \ref{lab_alg01} and the formula in Theorem \ref{lab_MainThorem_02x} (3). We found that numerical and theoretical results are identical.
   \begin{equation*}
   \begin{split}
   (p,q)&=\\
   &\lbrace (19, 83), (19, 443), (59, 43), (59, 83),\\
   &(59, 163), (59, 283), (59, 443), (139, 83),\\
   &(139,443),(179,43),(179,83),(179,163),\\
   &(179,283),(179,443),(379,83),(379,443),\\
   &(419, 43), (419, 83), (419, 163), (419, 283),\cdots\rbrace.
   \end{split}
   \end{equation*}
   \end{sec3Bexp1}
    For example, for all 
    \begin{equation*}
    \begin{split}
    (p,q)\in \lbrace (59, 43), (59, 83),(59, 163), (59, 283), (59, 443) \rbrace,
    \end{split}
    \end{equation*}
     $ G_{s}(\beta^{k}) $ has same number of zeros which equals $ \tfrac{p-1}{2}+2=31 $, where $ k\in Z_{2T} $, but for each pair of $ (p,q) $, the linear complexity is different, that is equal to $ 2pq- \tfrac{p-1}{2}-2=2pq-31 $. For instance, if $ p=59 $ and $ q=43 $, then $ LC(s) =5043$.
    \subsection{$ 2 $ is a quadratic non-residue  modulo $ p $ and a quadratic residue modulo $ q $}
    
    \begin{sec3lemma10}\label{lab_sec3_lemma10}
    Let $ 2\in  D_{1}^{(p)}$ and $  2\in  D_{0}^{(q)} $. For $ k\in Z_{2T} $, let $ k_{1}\equiv k\pmod p $ and $ k_{2}\equiv k\pmod q $. Then, the values of $ G_{s}(\beta^{k}) $ can be determined according to various range of $ k $.
    \begin{enumerate} 
    \item $ k\in \lbrace 0,\ pq \rbrace \cup 2qZ^{*}_{p}\cup pZ^{*}_{2 q} $.
    \begin{equation*}
    G_{s}(\beta^{k})=
    \begin{cases}
    3pq-1&k\in\lbrace 0\rbrace,\\
    2pq-4&k\in\lbrace pq\rbrace,\\
    -1   &k\in 2qZ^{*}_{p},\\
    -4   &k\in pZ^{*}_{2q}.
    \end{cases}
    \end{equation*}
    \item $ k\in D_{0}^{(2T)}\cup D_{1}^{(2T)} $.
    \begin{enumerate}
    \item
    If $ k\in D_{0}^{(2T)} $ and $ (k_{1},\ k_{2})\in D^{(p)}_{0}\times D^{(q)}_{0} $ and $ \biggl(\frac{q}{p}\biggr)=\biggl(\frac{p}{q}\biggr) $ or $ k\in D_{1}^{(2T)} $ and $ (k_{1},\ k_{2})\in D^{(p)}_{0}\times D^{(q)}_{1} $ and $ \biggl(\frac{q}{p}\biggr)=-\biggl(\frac{p}{q}\biggr) $, then
    \begin{equation*}
    G_{s}(\beta^{k})=(A_{0}-A_{1})(B_{0}-B_{1}+1)-4.
    \end{equation*} 
    \item
    If $ k\in D_{0}^{(2T)} $ and $ (k_{1},\ k_{2})\in D^{(p)}_{1}\times D^{(q)}_{1} $ and $ \biggl(\frac{q}{p}\biggr)=\biggl(\frac{p}{q}\biggr) $ or $ k\in D_{1}^{(2T)} $ and $ (k_{1},\ k_{2})\in D^{(p)}_{1}\times D^{(q)}_{0} $ and $ \biggl(\frac{q}{p}\biggr)=-\biggl(\frac{p}{q}\biggr) $, then
    \begin{equation*}
    G_{s}(\beta^{k})=(A_{0}-A_{1})(B_{0}-B_{1}-1)-4.
    \end{equation*} 
    \item
    If $ k\in D_{1}^{(2T)} $ and $ (k_{1},\ k_{2})\in D^{(p)}_{0}\times D^{(q)}_{1} $ and $ \biggl(\frac{q}{p}\biggr)=\biggl(\frac{p}{q}\biggr) $ or $ k\in D_{0}^{(2T)} $ and $ (k_{1},\ k_{2})\in D^{(p)}_{0}\times D^{(q)}_{0} $ and $ \biggl(\frac{q}{p}\biggr)=-\biggl(\frac{p}{q}\biggr) $, then
    \begin{equation*}
    G_{s}(\beta^{k})=(A_{1}-A_{0})(B_{0}-B_{1}-1)-4.
    \end{equation*} 
    \item
    If $ k\in D_{1}^{(2T)} $ and $ (k_{1},\ k_{2})\in D^{(p)}_{1}\times D^{(q)}_{0} $ and $ \biggl(\frac{q}{p}\biggr)=\biggl(\frac{p}{q}\biggr) $ or $ k\in D_{0}^{(2T)} $ and $ (k_{1},\ k_{2})\in D^{(p)}_{1}\times D^{(q)}_{1} $ and $ \biggl(\frac{q}{p}\biggr)=-\biggl(\frac{p}{q}\biggr) $, then
    \begin{equation*}
    G_{s}(\beta^{k})=(A_{1}-A_{0})(B_{0}-B_{1}+1)-4.
    \end{equation*}   
    \end{enumerate}
    \item $ k\in 2D_{0}^{(T)}\cup 2D_{1}^{(T)} $.
    \begin{enumerate}
    \item
    If $ k\in 2Z_{T}^{*} $ and $ (k_{1},\ k_{2})\in D^{(p)}_{1}\times D^{(q)}_{1}\cup  D^{(p)}_{0}\times D^{(q)}_{1}$ and $ \biggl(\frac{q}{p}\biggr)=\biggl(\frac{p}{q}\biggr) $ or $ k\in 2Z_{T}^{*} $ and $ (k_{1},\ k_{2})\in D^{(p)}_{1}\times D^{(q)}_{1}\cup  D^{(p)}_{0}\times D^{(q)}_{1}$ and $ \biggl(\frac{q}{p}\biggr)=-\biggl(\frac{p}{q}\biggr) $, then
    \begin{equation*}
    G_{s}(\beta^{k})=2B_{0}.
    \end{equation*} 
    \item
    If $ k\in 2Z_{T}^{*} $ and $ (k_{1},\ k_{2})\in D^{(p)}_{0}\times D^{(q)}_{0}\cup  D^{(p)}_{1}\times D^{(q)}_{0}$ and $ \biggl(\frac{q}{p}\biggr)=\biggl(\frac{p}{q}\biggr) $ or $ k\in 2Z_{T}^{*} $ and $ (k_{1},\ k_{2})\in D^{(p)}_{0}\times D^{(q)}_{0}\cup  D^{(p)}_{1}\times D^{(q)}_{0}$ and $ \biggl(\frac{q}{p}\biggr)=-\biggl(\frac{p}{q}\biggr) $, then
    \begin{equation*}
    G_{s}(\beta^{k})=2B_{1}.
    \end{equation*}  
    \end{enumerate}
    \item $ k\in qZ^{*}_{2p} $.
    \begin{enumerate}
    \item
    If  $ k_{1}\in D_{1}^{(p)} $ and $ \biggl(\frac{q}{p}\biggr)=\biggl(\frac{p}{q}\biggr) $ and $ \biggl(\frac{q}{p}\biggr)=1 $ or  $ k_{1}\in D_{0}^{(p)} $ and $ \biggl(\frac{q}{p}\biggr)=\biggl(\frac{p}{q}\biggr) $ and $ \biggl(\frac{q}{p}\biggr)=-1 $ or  $ k_{1}\in D_{1}^{(p)} $ and $ \biggl(\frac{q}{p}\biggr)=-\biggl(\frac{p}{q}\biggr) $ and $ \biggl(\frac{q}{p}\biggr)=1 $ or  $ k_{1}\in D_{0}^{(p)} $ and $ \biggl(\frac{q}{p}\biggr)=-\biggl(\frac{p}{q}\biggr) $ and $ \biggl(\frac{q}{p}\biggr)=-1 $, then
    \begin{equation*}
    G_{s}(\beta^{k})=A_{0}-A_{1}-4.
    \end{equation*}
    \item
    If  $ k_{1}\in D_{0}^{(p)} $ and $ \biggl(\frac{q}{p}\biggr)=\biggl(\frac{p}{q}\biggr) $ and $ \biggl(\frac{q}{p}\biggr)=1 $ or $ k_{1}\in D_{1}^{(p)} $ and $ \biggl(\frac{q}{p}\biggr)=\biggl(\frac{p}{q}\biggr) $ and $ \biggl(\frac{q}{p}\biggr)=-1 $ or  $ k_{1}\in D_{0}^{(p)} $ and $ \biggl(\frac{q}{p}\biggr)=-\biggl(\frac{p}{q}\biggr) $ and $ \biggl(\frac{q}{p}\biggr)=1 $ or $ k_{1}\in D_{1}^{(p)} $ and $ \biggl(\frac{q}{p}\biggr)=-\biggl(\frac{p}{q}\biggr) $ and $ \biggl(\frac{q}{p}\biggr)=-1 $, then
    \begin{equation*}
    G_{s}(\beta^{k})=A_{1}-A_{0}-4.
    \end{equation*}
    \end{enumerate}
     \item $ k\in 2pZ^{*}_{q} $.
       \begin{enumerate}
       \item
       If  $ k_{2}\in D_{1}^{(q)} $ and $ \biggl(\frac{q}{p}\biggr)=\biggl(\frac{p}{q}\biggr) $ and $ \biggl(\frac{p}{q}\biggr)=1 $ or  $ k_{2}\in D_{0}^{(q)} $ and $ \biggl(\frac{q}{p}\biggr)=\biggl(\frac{p}{q}\biggr) $ and $ \biggl(\frac{p}{q}\biggr)=-1 $ or  $ k_{2}\in D_{1}^{(q)} $ and $ \biggl(\frac{q}{p}\biggr)=-\biggl(\frac{p}{q}\biggr) $ and $ \biggl(\frac{p}{q}\biggr)=1 $ or  $ k_{2}\in D_{0}^{(q)} $ and $ \biggl(\frac{q}{p}\biggr)=-\biggl(\frac{p}{q}\biggr) $ and $ \biggl(\frac{p}{q}\biggr)=-1 $, then
       \begin{equation*}
       G_{s}(\beta^{k})=2B_{0}.
       \end{equation*}
       \item
       If  $ k_{2}\in D_{0}^{(q)} $ and $ \biggl(\frac{q}{p}\biggr)=\biggl(\frac{p}{q}\biggr) $ and $ \biggl(\frac{p}{q}\biggr)=1 $ or  $ k_{2}\in D_{1}^{(q)} $ and $ \biggl(\frac{q}{p}\biggr)=\biggl(\frac{p}{q}\biggr) $ and $ \biggl(\frac{p}{q}\biggr)=-1 $ or  $ k_{2}\in D_{0}^{(q)} $ and $ \biggl(\frac{q}{p}\biggr)=-\biggl(\frac{p}{q}\biggr) $ and $ \biggl(\frac{p}{q}\biggr)=1 $ or  $ k_{2}\in D_{1}^{(q)} $ and $ \biggl(\frac{q}{p}\biggr)=-\biggl(\frac{p}{q}\biggr) $ and $ \biggl(\frac{p}{q}\biggr)=-1 $, then
       \begin{equation*}
       G_{s}(\beta^{k})=2B_{1}.
       \end{equation*}
       \end{enumerate}
    \end{enumerate}
    \end{sec3lemma10}
    \begin{proof}
    Similar to the proof of Lemma \ref{lab_sec3_lemma8}.
    \end{proof} 
    \begin{sec3Ccor1}\label{lab_sec3Ccor1}
    Let $ 2\in D_{1}^{(p)} $, $ 2\in D_{0}^{(q)} $, $ \bigl(\tfrac{r}{p}\bigr) =1$ and $ \bigl(\tfrac{r}{q}\bigr) =1$. In addition, let  $ r\mid p\pm 4 $ if $ p\equiv\pm 3\pmod 8 $, and $ r\mid q\pm 1 $ if $ q\equiv\mp 1\pmod 8 $. Then, $ G_{s}(\beta^{k}) $ has exactly $ \tfrac{3(p-1)(q-1)}{4}+\tfrac{q-1}{2} $'s zeros, where $ k\in Z_{2T}^{*}\cup 2Z_{T}^{*}\cup 2pZ_{q}^{*} $.
    \end{sec3Ccor1}
    \begin{proof}
      As for the proof of Corollary \ref{lab_sec3Bcor1}, from the conditions given in the actual Lemma, we can obtain $ A_{0}-A_{1}=\pm 2 $ and $ B_{0},B_{1}\in \lbrace 0,-1\rbrace $ or $B_{0}-B_{1}=\pm 1  $. From Lemma \ref{lab_sec3_lemma10} (2), $ G_{s}(\beta^{k})=0 $ for $ k\in Z_{2T}^{*} $ in one of the four sub-cases, that contributes $ \tfrac{(p-1)(q-1)}{4} $ zeros. By the same way, we get $ \tfrac{(p-1)(q-1)}{2} $ zeros for $ k\in 2Z_{T}^{*} $ from Lemma \ref{lab_sec3_lemma10} (3), and $ \tfrac{q-1}{2} $ zeros for $ k\in 2pZ_{q}^{*} $ from Lemma \ref{lab_sec3_lemma10} (5), of $ G_{s}(\beta^{k}) $. Hence, the number of zeros of $ G_{s}(\beta^{k}) $ where $ k\in Z_{2T}^{*}\cup 2Z_{T}^{*}\cup 2pZ_{q}^{*} $ equals 
      \begin{equation*}
      \begin{split}   
      &\frac{(p-1)(q-1)}{4}+\frac{(p-1)(q-1)}{2}+\frac{q-1}{2}\\
      &=\frac{3(p-1)(q-1)}{4}+\frac{q-1}{2}.
      \end{split}
      \end{equation*}
    \end{proof} 
    \begin{sec3Ccor2}\label{lab_sec3Ccor2}
       Let $ 2\in D_{1}^{(p)} $, $ 2\in D_{0}^{(q)} $, $ \bigl(\tfrac{r}{p}\bigr) =1$ and $ \bigl(\tfrac{r}{q}\bigr) =1$. In addition, let  $ r\mid p\pm 16 $ if $ p\equiv\pm 3\pmod 8 $, and $ r\mid q\pm 1 $ if $ q\equiv\mp 1\pmod 8 $. Then, $ G_{s}(\beta^{k}) $ has exactly $ \tfrac{(p-1)(q-1)}{2}+\tfrac{p-1}{2}+\tfrac{q-1}{2} $'s zeros, where $ k\in qZ_{2p}^{*}\cup 2Z_{T}^{*}\cup 2pZ_{q}^{*} $.
    \end{sec3Ccor2}
    \begin{proof}
    By Lemma \ref{lab_sec2_lemma4x}, $ A_{0},A_{1},B_{0},B_{1}\in \mathbf{GF}(r) $. By Lemma \ref{lab_sec2_lemma1x} and referring to the proof of Corollary \ref{lab_sec3Bcor1}, we obtain $ A_{0}-A_{1}=\pm 4 $ and $ B_{0},B_{1}\in \lbrace 0,-1\rbrace $ or $B_{0}-B_{1}=\pm 1  $. Looking at Lemma \ref{lab_sec3_lemma10} (4) with $ A_{0}-A_{1}=\pm 4 $, we remark that $ G_{s}(\beta^{k})=0 $ for $ k\in qZ_{2p}^{*} $ in one of the two sub-cases, that contributes $ \tfrac{p-1}{2} $ zeros. On the other hand, the condition $ B_{0},B_{1}\in \lbrace 0,-1\rbrace $ leads to $ \tfrac{(p-1)(q-1)}{2} +\tfrac{q-1}{2}$ zeros as proved in Corollary \ref{lab_sec3Ccor1}. Hence, the number of zeros of $ G_{s}(\beta^{k}) $ where $ k\in qZ_{2p}^{*}\cup 2Z_{T}^{*}\cup 2pZ_{q}^{*} $ equals 
         \begin{equation*}     
         \frac{(p-1)(q-1)}{2}+\frac{p-1}{2}+\frac{q-1}{2}.
         \end{equation*} 
    \end{proof} 
    \begin{sec3Ccor3}\label{lab_sec3Ccor3}
    Let $ 2\in D_{1}^{(p)} $, $ 2\in D_{0}^{(q)} $, $ \bigl(\tfrac{r}{p}\bigr) =1$ and $ \bigl(\tfrac{r}{q}\bigr) =-1$. In addition, let  $ r\mid p\pm 16 $ if $ p\equiv\pm 3\pmod 8 $. Then, $ G_{s}(\beta^{k}) $ has exactly $ \tfrac{p-1}{2} $'s zeros, where $ k\in qZ_{2p}^{*} $.
    \end{sec3Ccor3}
    \begin{proof}
    Since $ \bigl(\tfrac{r}{p}\bigr) =1$ and $ \bigl(\tfrac{r}{q}\bigr) =-1$, by Lemma \ref{lab_sec2_lemma4x}, $ A_{0},A_{1}\in \mathbf{GF}(r) $ and $B_{0},B_{1}\in \mathbf{GF}(r^{m})\setminus \mathbf{GF}(r) $. The rest of proof is similar to that of Corollary \ref{lab_sec3Ccor2}.
    \end{proof} 
    \begin{sec3Ccor4}\label{lab_sec3Ccor4}
       Let $ 2\in D_{1}^{(p)} $, $ 2\in D_{0}^{(q)} $, $ \bigl(\tfrac{r}{p}\bigr) =-1$ and $ \bigl(\tfrac{r}{q}\bigr) =1$. In addition, let  $ r\mid q\pm 1 $ if $ q\equiv\mp 1\pmod 8 $. Then, $ G_{s}(\beta^{k}) $ has exactly $ \tfrac{(p-1)(q-1)}{2}+\tfrac{q-1}{2} $'s zeros, where $ k\in  2Z_{T}^{*}\cup 2pZ_{q}^{*} $.
       \end{sec3Ccor4}
     \begin{proof}
     See the proof for Corollaries \ref{lab_sec3Ccor1}-\ref{lab_sec3Ccor3}.
     \end{proof} 
     \begin{sec3Ccor5}\label{lab_sec3Ccor5}
     Let $ \bigl(\tfrac{r}{p}\bigr) =-1$ and $ \bigl(\tfrac{r}{q}\bigr) =-1$. Then, there are no $ r $ with $ r\geq 5 $ such that $ r\mid 3pq-1 $ and $ r\mid 2pq-4 $.
     \end{sec3Ccor5}
     \begin{proof}
     The conditions $ r\mid 3pq-1 $ and $ r\mid 2pq-4 $ imply that $ r=5 $. But it is impossible that $ r=5 $ occurs. If $ r=5 $, by the Quadratic Reciprocity Law in Lemma \ref{lab_sec2_lemma3x} (2), $ \bigl(\tfrac{p}{5}\bigr) =\bigl(\tfrac{q}{5}\bigr)=\bigl(\tfrac{5}{p}\bigr)=\bigl(\tfrac{5}{q}\bigr)=-1$.With  $ r\mid 3pq-1 $, it leads to $ 3pq\equiv 1\pmod 5 $. Hence, $ \bigl(\tfrac{3pq}{5}\bigr)=\bigl(\tfrac{3}{5}\bigr)\cdot\bigl(\tfrac{p}{5}\bigr) \cdot\bigl(\tfrac{q}{5}\bigr)=-1=\bigl(\tfrac{1}{5}\bigr)=1$, that is a contradiction.    
     \end{proof} 
     \begin{sec3CThm01}\label{lab_sec3CThm01}
    Let $ 2\in D_{1}^{(p)} $, $ 2\in D_{0}^{(q)} $, $ \bigl(\tfrac{r}{p}\bigr) =1$ and $ \bigl(\tfrac{r}{q}\bigr) =1$. In addition, let  $ r\mid p\pm 4 $ if $ p\equiv\pm 3\pmod 8 $, and $ r\mid q\pm 1 $ if $ q\equiv\mp 1\pmod 8 $.
    Then, the linear complexity over $ \mathbf{GF}(r^{m})$ of the generalized cyclotomic quaternary sequences with period $ 2T $ specified in (\ref{lab_seq_chang_ex}) can be determined according to following cases:
     \begin{enumerate} 
     \item if $ (p,q)\in (3,1)\cup (-3,-1) $ and $r\ne 13$ \textbf{OR} $ (p,q)\in (3,-1)$ and $ r\ne 11$ , then
     \begin{equation*}
     LC(s)=\frac{5pq+3p+q-1}{4}
     \end{equation*}
    
      \item If $ (p,q)\in (3,1)\cup (-3,-1) $ and $r= 13$ \textbf{OR} $ (p,q)\in (3,-1)$ and $ r= 11$, then
     \begin{equation*}
     LC(s)=\frac{5pq+3p+q-5}{4}
      \end{equation*}
      \end{enumerate}
     \end{sec3CThm01}
     \begin{sec3Cremark1}\label{Lab_sec3Cremark1}
     We consider two particular conditions that are $ r=13 $ for the case where $ (p,q)\in (3,1)\cup (-3,-1) $, and $ r=11 $ for the case where $ (p,q)\in (3,-1)$ in Theorem \ref{lab_sec3CThm01}, respectively. It is easy to check that, for the two cases if  $ r=13 $ and $ r=11 $ respectively, then $ r\mid 3pq-1 $. In other words, by Lemma \ref{lab_sec3_lemma9} (1), $ \beta^{0} =1$ is a zero for $ G_{s}(\beta^{k}) $ where $ k\in Z_{2T} $.
     \end{sec3Cremark1}
     \begin{proof}[Proof of Theorem \ref{lab_sec3CThm01}]
     Using Corollary \ref{lab_sec3Ccor1} and (\ref{lab_eq_GCD_LC_eq_ex}) to prove Theorem \ref{lab_sec3CThm01} (1), and  Corollary \ref{lab_sec3Ccor1}, (\ref{lab_eq_GCD_LC_eq_ex}), and Remark \ref{Lab_sec3Cremark1} to prove Theorem \ref{lab_sec3CThm01} (2).
     \end{proof}
      
       \begin{sec3CThm02}\label{lab_sec3CThm02}
         Let $ 2\in D_{1}^{(p)} $, $ 2\in D_{0}^{(q)} $, $ \bigl(\tfrac{r}{p}\bigr) =1$ and $ \bigl(\tfrac{r}{q}\bigr) =1$. In addition, let  $ r\mid p\pm 16 $ if $ p\equiv\pm 3\pmod 8 $, and $ r\mid q\pm 1 $ if $ q\equiv\mp 1\pmod 8 $.
         Then, the linear complexity over $ \mathbf{GF}(r^{m}) $ of the generalized cyclotomic quaternary sequences with period $ 2T $ specified in (\ref{lab_seq_chang_ex}) can be determined according to following cases: 
      \begin{enumerate} 
       \item if $ (p,q)\in (3,1)\cup (-3,-1) $ and  $r\ne 7$ \textbf{OR} $ (p,q)\in (3,-1) $ and $r\notin \left\lbrace 7,47\right\rbrace$ , then 
      \begin{equation*}
       LC(s)=\frac{3pq+1}{2}
       \end{equation*}
      
        \item If $ (p,q)\in (3,1)\cup (-3,-1) $ and  $r=7$ \textbf{OR} $ (p,q)\in (3,-1) $ and $r\in \left\lbrace 7,47\right\rbrace$ , then 
    \begin{equation*}
       LC(s)=\frac{3pq-1}{2}
       \end{equation*}
      
    \end{enumerate}
     \end{sec3CThm02}
     \begin{sec3Cremark2}\label{Lab_sec3Cremark2}
     In Theorem \ref{lab_sec3CThm02},  $r=7$ implies that $ \beta^{0}=1 $ is a zero of $ G_{s}(\beta^{k}) $ where $ k\in Z_{2T} $ for $ (p,q)\in (3,1)\cup (-3,-1) $. While for $ (p,q)\in (3,-1) $, $r=47$ implicates that $ \beta^{0}=1 $ is a zero, and $r=7$ implies that $ \beta^{pq}$ is a zero, of $ G_{s}(\beta^{k}) $ where $ k\in Z_{2T} $ .
     \end{sec3Cremark2}
     \begin{proof}[Proof of Theorem \ref{lab_sec3CThm02}]
         Using Corollary \ref{lab_sec3Ccor2} and (\ref{lab_eq_GCD_LC_eq_ex}) to prove Theorem \ref{lab_sec3CThm02} (1), and  Corollary \ref{lab_sec3Ccor2}, (\ref{lab_eq_GCD_LC_eq_ex}), and Remark \ref{Lab_sec3Cremark2} to prove Theorem \ref{lab_sec3CThm02} (2).
         \end{proof}
     
    \begin{sec3CThm03}\label{lab_sec3CThm03}
     Let $ 2\in D_{1}^{(p)} $, $ 2\in D_{0}^{(q)} $, $ \bigl(\tfrac{r}{p}\bigr) =1$ and $ \bigl(\tfrac{r}{q}\bigr) =-1$. In addition, let  $ r\mid p\pm 16 $ if $ p\equiv\pm 3\pmod 8 $. Then, the linear complexity over $\mathbf{GF}(r^{m}) $ of the generalized cyclotomic quaternary sequences with period $ 2T $ given in (\ref{lab_seq_chang_ex}) can be determined according to following cases:
     \begin{enumerate} 
    \item if $ (p,q)\in (3,\pm 1) $ and $ r\nmid 48q+1 $ and $ r\nmid 8q+1 $ \textbf{OR} $ (p,q)\in (-3,-1) $ and $ r\nmid 48q-1 $ and $ r\nmid 8q-1 $, then 
     \begin{equation*}
     LC(s)=\frac{4pq-p+1}{2}.
     \end{equation*}
     
       \item If $ (p,q)\in (3,\pm 1) $ and ($ r\mid 48q+1 $ or $ r\mid 8q+1 $) and $ r\ne 5 $ \textbf{OR} $ (p,q)\in (-3,-1) $ and ($ r\mid 48q-1 $ or $ r\mid 8q-1 $) and $ r\ne 5 $, then
      \begin{equation*}
       LC(s)=\frac{4pq-p-1}{2}.
       \end{equation*}
       
         \item If $ (p,q)\in (3,\pm 1) $  and ($ r\mid 48q+1 $ or $ r\mid 8q+1 $) and $ r=5 $ \textbf{OR} $ (p,q)\in (-3,-1) $ and ($ r\mid 48q-1 $ or $ r\mid 8q-1 $) and $ r=5 $, then
        \begin{equation*}
           LC(s)=\frac{4pq-p-3}{2}.
           \end{equation*}
       \end{enumerate}
        \end{sec3CThm03}
    \begin{sec3Cremark3}\label{Lab_sec3Cremark3}
    In Theorem \ref{lab_sec3CThm03}, for $ (p,q)\in (3,\pm 1) $, $ r\mid 48q+1 $ and $ r\mid 8q+1 $ imply that $ \beta^{0}=1 $ and $ \beta^{pq}$ are zeros of $ G_{s}(\beta^{k}) $ where $ k\in Z_{2T} $, respectively. While for $ (p,q)\in (-3,-1) $, $ \beta^{0}=1 $ is a zero of $ G_{s}(\beta^{k}) $ if only if $ r\mid 48q-1 $, and $ \beta^{pq}$ is a zero of $ G_{s}(\beta^{k}) $ if only if $ r\mid 8q-1 $. It is obvious that if $ r=5 $, then $ r\mid 48q\pm 1 $ implicate $ r\mid 8q\pm 1 $, vice versa.
    \end{sec3Cremark3}
    \begin{proof}[Proof of Theorem \ref{lab_sec3CThm03}]
            Using Corollary \ref{lab_sec3Ccor3} and (\ref{lab_eq_GCD_LC_eq_ex}) to prove Theorem \ref{lab_sec3CThm02} (1), and  Corollary \ref{lab_sec3Ccor3}, (\ref{lab_eq_GCD_LC_eq_ex}), and Remark \ref{Lab_sec3Cremark3} to prove Theorem \ref{lab_sec3CThm03} (2), and (3).
     \end{proof}
    \begin{sec3CThm04}\label{lab_sec3CThm04}
    Let $ 2\in D_{1}^{(p)} $, $ 2\in D_{0}^{(q)} $, $ \bigl(\tfrac{r}{p}\bigr) =-1$ and $ \bigl(\tfrac{r}{q}\bigr) =1$. In addition, let $ r\mid q\pm 1 $ if $ q\equiv\mp 1\pmod 8 $. Then, the linear complexity over $ \mathbf{GF}(r^{m}) $ of the generalized cyclotomic quaternary sequences with period $ 2T $ given in (\ref{lab_seq_chang_ex}) can be determined according to following cases: 
    \begin{enumerate}
     \item if $ (p,q)\in (3,1) $ and $ r\nmid 3p-1 $ and $ r\nmid 2p-4 $ \textbf{OR} $ (p,q)\in (\pm 3,-1) $  and $ r\nmid 3p+1 $ and $ r\nmid 2p+4 $, then 
           \begin{equation*}
           LC(s)=\frac{3pq+p}{2}.
           \end{equation*}
    \item If $ (p,q)\in (3,1) $ and ( $ r\mid 3p-1 $ or $ r\mid 2p-4 $) and $ r\ne 5 $ \textbf{OR} $ (p,q)\in (\pm 3,-1) $ and ( $ r\mid 3p+1 $ or $ r\mid 2p+4 $) and $ r\ne 5 $, then
             \begin{equation*}
                LC(s)=\frac{3pq+p-2}{2}.
                \end{equation*}
    \item If $ (p,q)\in (3,1) $  and ($ r\mid 3p-1 $ or $ r\mid 2p-4 $) and $ r=5 $ \textbf{OR } $ (p,q)\in (\pm 3,-1) $  and ($ r\mid 3p+1 $ or $ r\mid 2p+4 $) and $ r=5 $, then
                  \begin{equation*}
                  LC(s)=\frac{3pq+p-4}{2}.
                  \end{equation*}
            \end{enumerate}
    \end{sec3CThm04}
    \begin{sec3Cremark4}\label{Lab_sec3Cremark4}
    In Theorem \ref{lab_sec3CThm04}, the conditions $ r\mid 3p+1 $ and $ r\mid 2p+4 $ signify that  $ \beta^{0}=1 $ and $ \beta^{pq} $ are zeros of  $ G_{s}(\beta^{k}) $ for $ k\in Z_{2T} $ if $ (p,q)\in (\pm 3,-1) $, respectively. If $ (p,q)\in (3,1) $, the same conditions change into $ r\mid 3p-1 $ and $ r\mid 2p-4$, respectively. It is clear that if $ r=5 $, then $ r\mid 3p\pm 1 $ implicate $ r\mid 2p\pm 4 $, vice versa.
    \end{sec3Cremark4}
    \begin{proof}[Proof of Theorem \ref{lab_sec3CThm04}]
    Using Corollary \ref{lab_sec3Ccor4} and (\ref{lab_eq_GCD_LC_eq_ex}) to prove Theorem \ref{lab_sec3CThm04} (1), and  Corollary \ref{lab_sec3Ccor4}, (\ref{lab_eq_GCD_LC_eq_ex}), and Remark \ref{Lab_sec3Cremark4} to prove Theorem \ref{lab_sec3CThm04} (2), and (3).
    \end{proof}
    
    \begin{sec3CThm05}\label{lab_sec3CThm05}
     Let $ \bigl(\tfrac{r}{p}\bigr) =-1$ and $ \bigl(\tfrac{r}{q}\bigr) =-1$. Then, the linear complexity over $ \mathbf{GF}(r^{m}) $ of the generalized cyclotomic quaternary sequences with period $ 2T $ specified in (\ref{lab_seq_chang_ex}) can be determined according to following cases:
     \begin{enumerate}
     \item if $  r\nmid 3pq-1 $ and $ r\nmid 2pq-4 $, then 
      \begin{equation*}
       LC(s)=2pq.
     \end{equation*}
      \item If $  r\mid 3pq-1 $ or $ r\mid 2pq-4 $, then 
          \begin{equation*}
           LC(s)=2pq-1.
         \end{equation*}
     \end{enumerate}
        
    \end{sec3CThm05}
   \begin{proof}
   Since $ \bigl(\tfrac{r}{p}\bigr) =-1$ and $ \bigl(\tfrac{r}{q}\bigr) =-1$, by Lemma \ref{lab_sec2_lemma4x}, $ A_{0},A_{1},B_{0},B_{1}\in \mathbf{GF}(r^{m})\setminus \mathbf{GF}(r)$. From Lemma \ref{lab_sec3_lemma9}, $ G_{s}(\beta^{k}) $ has no zeros for $ k\in Z_{2T} $, except for two possible zeros which are $ \beta^{0} =1$ and $ \beta^{pq}$, respectively. By Corollary \ref{lab_sec3Ccor5}, no $ r $ such that both $ r\mid 3pq-1 $ and $ r\mid 2pq-4 $ hold together.
   \end{proof}

 \section{Conclusion}
 In this paper, we computed out the linear complexity over $ \mathbf{GF}(r) $  of the  generalized cyclotomic quaternary sequences of length $ 2T $ specified in (\ref{lab_seq_chang_ex}). The results show that the minimal value of the linear complexity  is equal to $ \tfrac{5pq+p+q+1}{4} $ which is greater than $ T $, the half of the period of the sequences. According to the Berlekamp-Massey algorithm \cite{B5}, these sequences are viewed as enough good for the use in cryptography. Remark that if the character of the extension field $ \mathbf{GF}(r^{m}) $, $ r $, is chosen so that $ r\geq 5 $, $ r\ne p,q $, $ \bigl(\tfrac{r}{p}\bigr)=\bigl(\tfrac{r}{q}\bigr)=-1 $, $ r\nmid 3pq-1 $, and $ r\nmid 2pq-4 $, then by Theorem \ref{lab_sec3CThm05}, the linear complexity can reach the maximal value which is the length of the sequences, $ 2T $.

\bibliographystyle{amsplain}

\begin{thebibliography}{99}
\bibitem{B1} 
M. K. Simon, J. K. Omura, R. A. Schotz,  B. K. Levitt, Spread Spectrum Communications, vol. 1, Rockville MD: Computer Science Press, 1985.
\bibitem{B2} 
I. B. Damgard,  On the randomness of Legendre and Jacobi sequences, Advances in Cryptograpy(CRYPTO’88) LNCS 403 (1990) 163-172, Springer-Verlag, Berlin.
\bibitem{B3} 
X. Yang, C. Ding,  New classes of balanced quaternary sequences and almost balanced binary sequences with optimal autocorrelation value, IEEE Trans. Inf. Theory 56 (12)  (2010) 6398-6405.

\bibitem{B4} 
C. Ding, Cyclic codes from cyclotomic sequences of order four, Finite Fields and Their Appl. 23 (2013) 8-24.

\bibitem{B5} 
J. L. Massey, Shift register synthesis and BCH decoding, IEEE Trans. Inf. Theory IT-15 (1) (1969) 122-127.

\bibitem{B6}
A. L. Whiteman,  A family of difference sets, Illinois J. Math. (1962) 107-121.

\bibitem{B7} 
C. Ding,  Linear comlexity of generalized cyclotomic binary sequence of order 2, Finite Fields and Their Applications 3 (1997) 159-174.

\bibitem{B8} 
C. Ding,  Autocorrelation values of generalized cyclotomic sequences of order two, IEEE Trans. Inf. Theory 44 (5) (1998) 1699-1702.

\bibitem{B9} 
C. Ding, T. Helleseth, W. Shan, On the linear complexity of Legendre sequences, IEEE Trans. Inf. Theory 44 (3) (1998) 1276-1278.

\bibitem{B10} 
X. Du, Z. Chen,  Linear complexity of quaternary sequences generated using generalized cyclotomic classes modulo $ 2p $, IEICE Trans. Fundamentals EA94-A (5) (2011) 1214-1217.
\bibitem{B11} 
Y. J. Kim, Y. P. Hong,  H. Y. Song, Autocorrelation of some quaternary cyclotomic sequences of length $ 2p $, IEICE Trans. Fundamentals E91-A (12) (2008) 3679-3684.

\bibitem{B12} 
P. Ke, Z. Yang,  J. Zhang, On the autocorrelation and linear complexity of some $ 2p $ periodic quaternary cyclotomic sequences over $ \mathbf{F}_{4} $, IEICE Trans. Fundamentals EA94-A (11) (2011) 2472-2477.

\bibitem{B13} 
Z. Chang,  D. Li,  On the linear complexity of quaternary cyclotomic sequences with the period $ 2pq $, IEICE Trans. Fundamentals EA97-A (2) (2014) 679-684.
\bibitem{B14} 
D. Li, Q. Wen, J. Zhang, Z. Chang, Linear complexity of generalized cyclotomic quaternary sequences with period $ pq $, IEICE Trans. Fundamentals E97-A (5)  (2014)  1153-1158.

\bibitem{B15} 
L. Zhao, Q. Wen,  J. Zhang,   On the linear complexity of a class of quaternary sequences with low autocorrelation, IEICE Trans. Fundamentals EA96-A (5) (2013) 997-1000.

\bibitem{B16} 
J. Zhang, C. Zhao,  X. Ma,  On the linear complexity of generalized cyclotomic binary sequences with length $ 2p^{2} $, IEICE Trans. Fundamentals EA93-A (1) (2010) 302-308.

\bibitem{B17} 
X. Du,  Z. Chen,  Trace Representation of Binary Generalized Cyclotomic Sequences with Length $ p^{m} $, IEICE Trans. Fundamantals E94-A (2) (2011) 761-765.

\bibitem{B18} 
Q. Wang, D. Lin,  X. Guang, On the linear complexity of Legendre sequences over $ \mathbb{F}_{q} $, IEICE Trans. Fundamantals E97-A (7) ( 2014) 1627-1630.

\bibitem{B19} 
D. M. Burton, Elementary Number Theory, fourth ed., McGraw-Hill International Editions, 1998.

\bibitem{B20}
L. Q. Hu, Q. Yue,  M. H. Wang, The linear complexity of Whiteman\textquoteright s generalized cyclotomic binary sequences with the period $ p^{m+1}q^{n+1} $, IEEE Trans. Inf. Theory 58 (8) (2012) 5534-5543.

\bibitem{B21} 
D. Li, Q. Wen, J. Zhang,  L. Jiang, Linear complexity of generalized cyclotomic binary sequences with period $ 2p^{m+1}q^{n+1} $, IEICE Trans. Fundamentals E98-A (6) (2015) 1244-1254.
\end{thebibliography}

\end{document}